\documentclass[]{aa}  

\usepackage{graphicx}
\usepackage{txfonts}
\usepackage{natbib}
\bibpunct{(}{)}{;}{a}{}{,} 

\usepackage{xcolor}
\begin{document} 

	\title{From gas to stars: MUSEings on the internal evolution of IC~1613}
	\subtitle{}
	
   \author{
        S.~Taibi\inst{1}
        \and
        G.~Battaglia\inst{2,3}
        \and
        M.~M.~Roth\inst{1,4}
        \and
        S.~Kamann\inst{5}
        \and
        G.~Iorio\inst{6,7,8}
        \and
        C.~Gallart\inst{2,3}
        \and
        R.~Leaman\inst{9}
        \and
        E.~D.~Skillman\inst{10}
        \and
        N.~Kacharov\inst{1}
        \and
        M.~A.~Beasley\inst{11,2,3}
	\and
        P.~E.~Mancera Piña\inst{12}
	\and
        G.~van de Ven\inst{9}
        }

   \institute{
            Leibniz-Institut für Astrophysik Potsdam (AIP), An der Sternwarte 16, D-14482 Potsdam, Germany\\
            \email{staibi@aip.de}
         \and
             Instituto de Astrofísica de Canarias, Calle Vía Láctea s/n, E-38206 La Laguna, Tenerife, Spain
        \and
            Universidad de La Laguna, Avda. Astrofísico Fco. Sánchez, E-38205 La Laguna, Tenerife, Spain
        \and
            Universit\"at Potsdam, Institut für Physik und Astronomie, Karl-Liebknecht-Str. 24/25, D-14476 Potsdam, Germany
	\and  
            Astrophysics Research Institute, Liverpool John Moores University, 146 Brownlow Hill, Liverpool L3 5RF, UK
        \and
            Dipartimento di Fisica e Astronomia “Galileo Galilei”, Università di Padova, vicolo dell’Osservatorio 3, IT-35122, Padova, Italy
        \and
            INAF - Osservatorio Astronomico di Padova, vicolo dell’Osservatorio 5, IT-35122 Padova, Italy
        \and
            INFN - Padova, Via Marzolo 8, I–35131 Padova, Italy
        \and
            Department of Astrophysics, University of Vienna, Türkenschanzstraße 17, 1180 Vienna, Austria
        \and
            Minnesota Institute for Astrophysics, University of Minnesota, 116 Church St.~SE, Minneapolis, MN 55455, USA
        \and
            Centre for Astrophysics and Supercomputing, Swinburne University, John Street, Hawthorn VIC 3122, Australia
	\and
            Leiden Observatory, Leiden University, P.O. Box 9513, 2300 RA, Leiden, The Netherlands
            }

   \date{Received; accepted}

 
  \abstract
   {The kinematics and chemical composition of stellar populations of different ages provide crucial information on the evolution of the various components of a galaxy.}
   {Our aim is to determine the kinematics of individual stars as a function of age in IC~1613, a star-forming, gas-rich and isolated dwarf galaxy of the Local Group (LG).}
   {We present results of a new spectroscopic survey of IC~1613 conducted with MUSE, an integral field spectrograph mounted on the Very Large Telescope. We extracted $\sim2000$ sources, from which we separated stellar objects for their subsequent spectral analysis. The quality of the dataset allowed us to obtain accurate classifications ($T_{\rm eff}$ to better than 500~K) and line-of-sight velocities (with average $\delta_v\sim7$~km\,s$^{-1}$) for about 800 stars. Our sample includes not only red giant branch (RGB) and main sequence (MS) stars, but also a number of probable Be and C stars. We also obtained reliable metallicities ($\delta_{\rm [Fe/H]}\sim0.25$~dex) for about 300 RGB stars.}
   {The kinematic analysis of IC~1613 revealed for the first time the presence of stellar rotation with high significance. We found general agreement with the rotation velocity of the neutral gas component. Examining the kinematics of stars as a function of broad age ranges, we find that the velocity dispersion increases as a function of age, with the behaviour being very clear in the outermost pointings, while the rotation-to-velocity dispersion support decreases. On timescales of $< 1$~Gyr, the stellar kinematics still follow very closely that of the neutral gas, while the two components decouple on longer timescales. The chemical analysis of the RGB stars revealed average properties comparable to other Local Group dwarf galaxies. We also provide a new estimation of the inclination angle using only independent stellar tracers.}
  {Our work provides the largest spectroscopic sample of an isolated LG dwarf galaxy. The results obtained seem to support the scenario in which the stars of a dwarf galaxy are born from a less turbulent gas over time.}


   \maketitle 


\section{Introduction}
\label{sec:intro}

The ability to resolve stellar populations in nearby galaxies allows us to study the physical processes that govern their evolution.
The dwarf galaxies of the Local Group (LG), in particular, provide an excellent laboratory due to their diverse characteristics, including size, gas content, star formation history (SFH), chemical abundances, and kinematics \citep[e.g.][]{Tolstoy2009,McConnachie2012,Kirby2013,Gallart2015,Simon2019,Putman2021,Battaglia+Nipoti2022}. Since these characteristics can be influenced by the proximity of a larger galaxy, such as the Milky Way and Andromeda, isolated LG dwarf galaxies provide valuable insights into the intrinsic mechanisms that regulate their evolution. 
In the gas-rich ones, the interplay between gravitational instabilities and stellar feedback processes can be reconstructed by combining the kinematics and spatial distribution of stars of different ages with the SFH and gas content of the galaxy \citep[e.g.][]{Leaman2017,Collins+Read2022}. Gathering this kind of information is usually an observational challenge, but the advent of integral field spectroscopy (IFS) has been transformative in this respect, due to the ability to inspect different galactic tracers with high observational efficiency (\citealp{Roth2018}, \citealp[but also][]{Sanchez2012,Cortese2014,Bundy2015,McLeod2020,Zoutendijk2020,Julio2023,Vaz2023}).

Among LG dwarf galaxies, IC~1613 is an ideal IFS target. On the one hand, it is a typical low-mass star-forming dwarf irregular galaxy \citep{Skillman2014}. On the other hand, stellar feedback processes have generated voids and bubbles in the interstellar medium whose impact on stellar and gas kinematics needs to be explored \citep{Read2016}.
First discovered by \citet{Wolf1906}, IC~1613 was recognised afterwards as an extragalactic object by Baade (1935, as reviewed by \citealp{Sandage1971}), who first determined its distance using Cepheid variable stars. Since then, several other measurements of its distance have been made using different indicators, including RR-Lyrae variable stars and the tip of the red giant branch (RGB). A compilation of literature distance values (including their own) was reported by \citet{Bernard2010}, who obtained a statistical average value of $(m-M)_0 = 24.400\pm0.014$, or $759\pm5$~kpc. 

IC~1613 therefore lies well beyond the Milky Way's (MW) virial radius, as well as that of M31 (being at $\sim520$~kpc distance from it, \citealp{McConnachie2012}). The systemic proper motion of IC~1613 obtained from \citet{McConnachie2021} with Gaia DR2 data does not rule out the possibility that the galaxy was ever within 300~kpc of M31, for M31 virial masses of $\gtrsim 1.3\times 10^{12}~M_\odot$. However, the most recent estimate of the systemic proper motion of IC~1613 by \citet{Bennet2023arXiv}, combining data from Gaia eDR3 and the Hubble Space Telescope (HST), makes such an association unlikely when assuming an M31 virial mass of $2\times 10^{12}~M_\odot$. 
Interestingly, \citet{Buck2019} assign to IC~1613 a high probability of being a backsplash galaxy (i.e. a currently isolated system that may have once passed close to a large host, but without becoming bound) of M31.

Its low luminosity (an absolute magnitude of $M_V =-15.2$~mag) corresponds to a total stellar mass of $M_* \sim10^8~M_\odot$, similar to that of other LG dIrrs like NGC~6822 and WLM \citep{McConnachie2012}. The SFH and stellar content of IC~1613 have been studied in detail over the years \citep[see e.g.][]{Cole1999,Skillman2003,Bernard2007,Skillman2014}. 
Based on deep HST Advanced Camera for Surveys (ACS) imaging, and after a review of archival HST data, \citet{Skillman2014} concluded that the SFH of IC~1613 has been close to constant on average throughout its life, with no evidence of an early dominant episode of star formation. We note that the HST/ACS field analysed by \citet{Skillman2014}, roughly located at the half-light radius of the galaxy, was considered as representative of the global SFH of IC~1613. The comparison with archival HST observations located at different radii supported this assumption.  
The average metallicity was linearly increasing through time, with values ranging from ${\rm [Fe/H]}\sim-2.0$~dex to $-0.8$~dex. 
Stars were formed at an average rate of $\psi(t) = 0.081\pm0.001~M_\odot {\rm yr}^{-1}$.

The structural properties of IC~1613 have been determined by studying the spatial distribution of stellar tracers in different evolutionary phases \citep[see, e.g.,][]{Albert2000,Borissova2000,Bernard2007,Garcia2009,Sibbons2015,McQuinn2017,Pucha2019,Higgs2021}. In general, it was found that the young stars are more centrally concentrated than the intermediate-age and old stars, a feature that is common in dwarfs. 
However, \citet{Pucha2019} using deep and wide Subaru/Hyper-SuprimeCam observations of IC~1613, showed that its young main sequence (MS) stars, along with the RGB and ancient horizontal branch (HB) stars, all extend to the outskirts of the galaxy, up to $\sim 24$~arcmin (i.e., $\sim4$ effective radii $R_e$). In particular, the young stars are found well beyond the currently active star forming regions (within $\sim1.5 \times R_e$), although with a much lower density compared to the intermediate and old age components. They also showed a steeper radial density profile in the inner regions than in the outer ones, in contrast to the RGB and HB stars. This seems to imply a different formation channel between the younger (e.g., from gas pushed outward through stellar feedback) and older stars (probably via gas accretion) in the galaxy's outskirts. In this work, we assume the structural parameters from \citet{Higgs2021} who conducted an homogeneous analysis of the isolated LG dwarf galaxies.

As other dIrrs, IC~1613 contains an extended component of neutral hydrogen (HI) gas, with a clumpy distribution on the inside, but showing regular contours at larger radii \citep[see][and references therein]{vanBergh2000}. In particular, the HI distribution is rich of shells and voids around the currently active star forming regions, where also the ionized gas is distributed \citep[e.g.,][]{Lozinskaya2003,Silich2006,Moiseev2012,Pokhrel2020,Yarovova2024}. The HI kinematics shows a linearly increasing rotation curve with a $v_{\rm max}\sim20$~km\,s$^{-1}$ \citep{Lake+Skillman1989,Oh2015,Read2016}. 
The rotation value, however, may represent a lower limit due to the fact that this galaxy is probably seen close to face-on. This translates into a significant uncertainty in the determination of the dynamical mass being between $M_{\rm dyn}\sim 5\times10^8 M_\odot$ and $\sim 8\times10^9 M_\odot$, depending on the assumed inclination angle \citep{Read2016}.

The first spectroscopic measurements of stars in IC~1613 were obtained from some of the brightest ones, that is, from 9 early B-type young supergiants \citep{Bresolin2007} and 3 evolved M-type supergiants \citep{Tautvaisiene2007}. Both studies obtained compatible metallicity values, $12+{\rm log(O/H)}=7.90\pm0.08$~dex and ${\rm [Fe/H]}=-0.67\pm0.09$~dex, respectively\footnote{Assuming ${\rm [O/Fe]}=-1$~dex and a solar oxygen abundance of $12+{\rm log(O/H)}=8.69$~dex \citep{Asplund2009}.}, later confirmed by \citet{Berger2018} who studied 21 young BA-type supergiant stars. Interestingly, these authors found a bimodal metallicity distribution that appears to be correlated with their spatial location (i.e. lower metallicity stars are found in high density HI regions). A comparable bi-modal distribution was also observed by \citet{Chun2022} for a sample of 14 red supergiants.
Furthermore, there is evidence from the youngest stellar population and evolved red supergiants of IC~1613 that the present-day ${\rm [\alpha/Fe]}$ ratio could be sub-solar (i.e., $\sim-0.1$~dex, \citealp{Tautvaisiene2007} and \citealp{Garcia2014}).
The chemical abundance of the interstellar medium has been obtained spectroscopically from the central HII regions and is in good agreement with results from the supergiant stars (\citealp{Bresolin2007}, but see also \citealp{Lee2003} and \citealp{Tautvaisiene2007}).

\citet{Kirby2013} were the first to present a statistically significant spectroscopic analysis of a sample of 125 RGB stars in IC~1613, which led to an average metallicity of ${\rm [Fe/H]}=-1.19\pm0.01$~dex, in agreement with the fact that older stars are in general more metal-poor than the younger ones, and with expectations from the age-metallicity relation found by \citet{Skillman2014}. On the same sample, \citet{Kirby2014} conducted a kinematic analysis which led to a systemic velocity of $-231.6\pm1.2$~km\,s$^{-1}$, in agreement with values from the HI component \citep{Lake+Skillman1989,Oh2015}, and a velocity dispersion of $10.8^{+1.0}_{-0.9}$~km\,s$^{-1}$. They did not find signs of rotation for the stellar component; this was later confirmed also by \citet{Wheeler2017} who re-analysed the \citeauthor{Kirby2013} dataset. This is not surprising considering that the spatial distribution of this dataset is roughly perpendicular to the HI major kinematic axis \citep{Lake+Skillman1989,Oh2015}, assuming that the stellar rotation follows that of neutral gas. 
The dynamical mass at the half-light radius reported by \citet{Kirby2014} was $M_{1/2}=1.1\pm0.2\times10^8 M_\odot$.

In this paper, we present a study of the stellar kinematic and chemical properties of IC~1613 from spectroscopic data taken with the MUSE integral field spectrograph on the Very Large Telescope (VLT). 
We took advantage of the unique capabilities of this instrument, that combines high spatial resolution with a 1 arcmin$^2$ field-of-view and a wide wavelength range at medium spectral resolution \citep{Bacon2014}.
We then performed a kinematic analysis as a function of stellar age and obtained metallicities for the largest spectroscopic stellar sample obtained to date for this galaxy.
IC~1613 is located at high Galactic latitude ($b \sim 61$~deg), so its low values for both foreground \citep{Schlafly+Finkbeiner2011} and internal reddening \citep{Georgiev1999} make it an ideal laboratory to compare the observed properties of its diverse stellar content.

The article is structured as follows: in Sect.~\ref{sec:observ} we give details on the data acquisition and reduction processes we conducted. Section~\ref{sec:velocity} is dedicated to spectral classification and velocity determination. In Sect.~\ref{sec:kinematics} we report details on the determination of likely member stars and their subsequent kinematic analysis, while in Sect.~\ref{sec:metallicity} we show results of the chemical analysis of the RGB stars in our sample. Section~\ref{sec:discussion} is dedicated to the discussion of the age-kinematic trends of the different tracers inspected in this work. In this section, we also discuss the dynamical mass estimation and the role played by the galaxy's inclination angle. Finally, Sect.~\ref{sec:conclusions} is dedicated to the summary and perspectives of future work, while in the appendices we report details on the sanity checks conducted during our analysis.
The parameters adopted for IC~1613 throughout the text are summarised in Table~\ref{table:1}.

\begin{table}
    \caption{Parameters adopted for IC~1613.}
    \label{table:1}      
    \centering          
    \begin{tabular}{l c c c}    
        \hline\hline
        Parameter & Units & Value  & Ref.\\
        \hline           
        $\alpha_{\rm J2000}$ & & $01^h04^m47.8^s$ & (1) \\
        $\delta_{\rm J2000}$ & & $+02^{\circ}07'04''$ & (1) \\
        $L_{\rm V}$ & $10^6 L_\odot$ & $100\pm15$ & (1) \\
        $D_\odot$ & kpc & $759\pm5$ & (2) \\
        $R_{\rm e}$ & arcmin & $6.8\pm0.05$  & (3) \\
                    & kpc    & $1.50\pm0.02$ & (3) \\
        P.~A. & deg & $90\pm1$ & (3) \\
        $\epsilon$ & & 0.20$\pm$0.05 & (3) \\
        & & & \\
        $\alpha_{\rm J2000, HI}$ & & $01^h04^m50.2^s$ & (4) \\
        $\delta_{\rm J2000, HI}$ & & $+02^{\circ}08'26''$ & (4) \\
        P.~A.$_{\rm HI}$ & deg & $72\pm2$ & (4) \\
        $i_{\rm HI}$     & deg & $39\pm2$ & (4) \\
        & & & \\
        $\bar{v}_{\rm sys}$ & km\,s$^{-1}$& $-231\pm1$ & (5)\\
        $\sigma_{\rm v}$ & km\,s$^{-1}$& $11.2\pm0.4$ & (5) \\
        $\left \langle \left [ {\rm Fe/H} \right ] \right \rangle$ & dex & $-1.06$ & (5)\\
        $\sigma_{\rm [Fe/H]}$ & dex & 0.26 & (5)\\
        \\
        Conversion & pc\,arcmin$^{-1}$ & 221 & \\
        \hline        
    \end{tabular}
    \tablefoot{The table lists: the coordinates of the galaxy's optical centre; the stellar luminosity in \textit{V}-band; the heliocentric distance; the effective radius ($R_{\rm e}=R_{\rm h} \sqrt{1-\epsilon}$); the position angle measured from north to east; the ellipticity ($\epsilon=1-b/a$); central coordinates, position angle and inclination of the HI velocity field; the chemo-kinematic parameters obtained in this work, i.e., the systemic velocity, the velocity dispersion, the median metallicity, and the intrinsic metallicity scatter.\\
    \textbf{References.} (1) \citet{McConnachie2012}; (2) \citet{Bernard2010}; (3) \citet{Higgs2021}; (4) \citet{Read2016}; (5) this work.
    }
\end{table}

\section{MUSE data of IC~1613}
\label{sec:observ}

\subsection{Observations and data reduction}
\label{subsec:reduction}

\begin{figure*}
	\includegraphics[width=\textwidth]{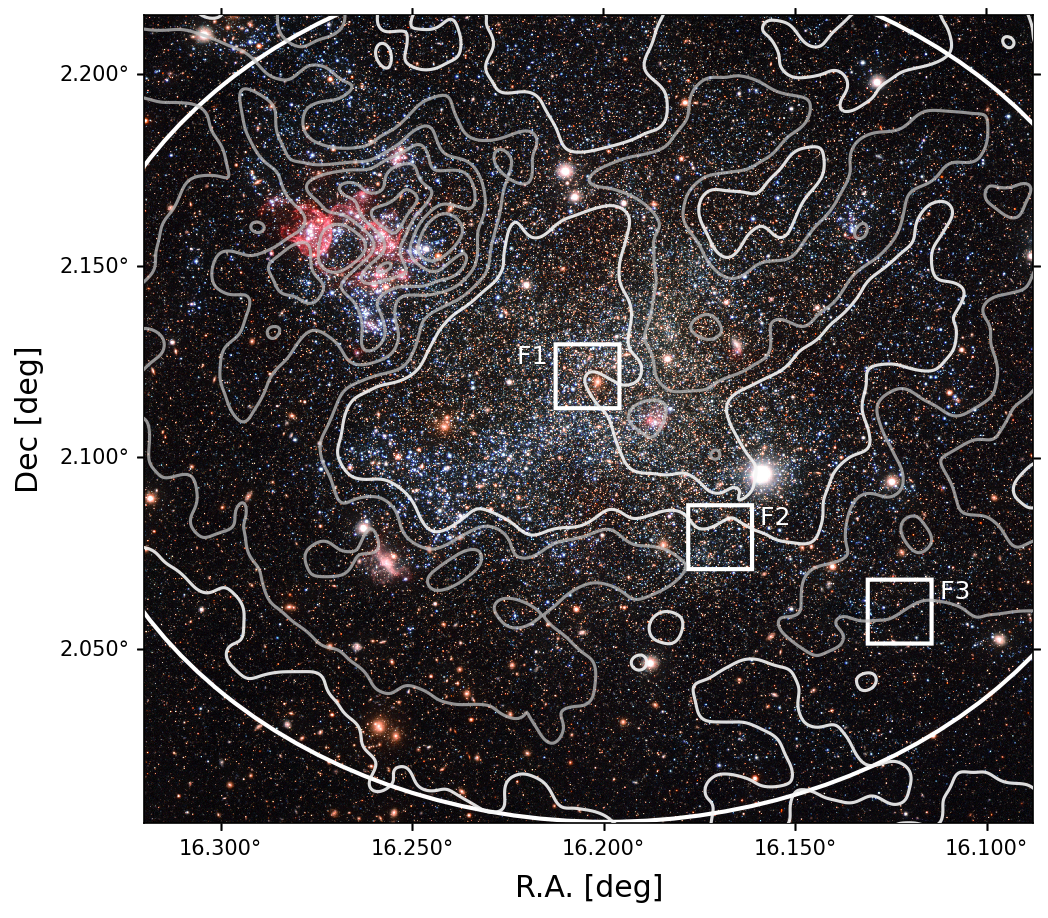}
	\caption{Finding chart showing the location of the 3 MUSE fields, marked as white boxes, overlaid over a wide-field image of IC~1613 (Credit: ESO - VST/Omegacam Local Group Survey). The white ellipse marks the half-light radius. The smoothed HI column density map from the Little-THINGS survey \citep{Hunter2012} is marked with logarithmically spaced isodensity contours (silver lines) starting at 3-$\sigma$ (white-smoke lines). North is up, East to the left}
	\label{fig:fov}
\end{figure*}

The data were acquired with VLT/MUSE in service mode\footnote{Under ESO programme 097.B-0373; PI: Battaglia.}.
MUSE is an integral field spectrograph with a spatial sampling of 0.2~arcsec and a spectral resolution varying between $R=1500-3000$ along its wavelength coverage. 
We used the instrument in nominal wide field mode, with a field of view of $\sim1\arcmin\times1\arcmin$ and a wavelength coverage between $4800 - 9300$~\AA.
We observed 3 fields (hereafter F1, F2, F3, moving outward from the centre; see Fig.~\ref{fig:fov} for their location) approximately along the projected semi-major axis of IC~1613, on the West side of the galaxy. We purposely avoided regions of on-going star formation.

Each field was observed with a total exposure time of 11080~s on source, split in $4\times2770$~s exposures, obtained by rotating the position angle of the spectrograph by 90$^{\circ}$ with respect to the previous exposure. No separate sky field was acquired since the stellar component of IC~1613 is resolved in the magnitude range of interest and the sky contribution is taken care of during the de-blending and spectral extraction phase of the data analysis (see Sect.~\ref{subsec:extraction}).
The data were acquired in dark time, clear sky, with a request for a seeing in V-band at zenith $\le0.9$~arcsec, fulfilled for all but one exposure of F1 (for which it deviated within 10\% from the request). The observing log is reported in Table~\ref{table:obs_log}.

\begin{figure*}
	\includegraphics[width=\textwidth]{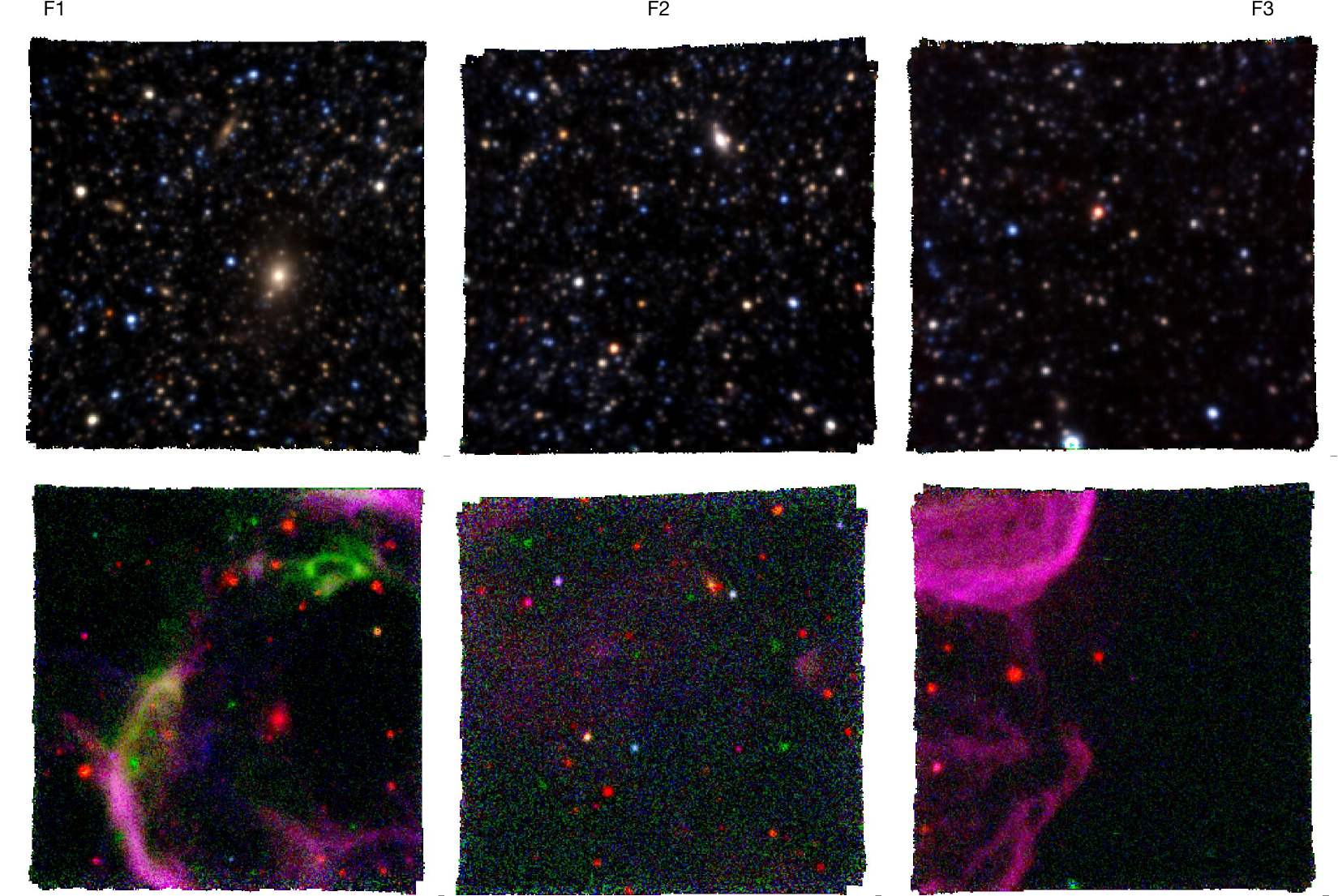}
	\caption{Broad- and narrow-band images obtained from the MUSE data cubes. \textit{Top row:} combined images in the \textit{VRI}-bands highlighting stellar sources and background galaxies. \textit{Bottom row:} combined images highlighting the emission-line ionised gas with H$_\alpha$ coloured in red, [SII] (6713~\AA) in blue, and [OIII] (5007~\AA) in green. In all panels, North is up and East is to the left.}
	\label{fig:rgbfig}
\end{figure*}

All data cubes were reduced using version 1.6 of the official MUSE pipeline \citep{Weilbacher2012}.
The basic steps of the reduction cascade (bias subtraction, slice tracing, wavelength calibration, and basic processing) were performed using the default settings\footnote{This included correcting for the observer's line of sight velocity with respect to the barycentre of the solar system.}.
This resulted in 24 pixel tables (one for each unit spectrograph) for each individual exposure.
When combining the pixel tables for each exposure, flux calibration was also applied using standard stars observed on the same nights.
Furthermore, the subtraction of sky emission lines was performed by determining their intensity directly from the scientific data, which contained sufficiently large patches of (almost) blank sky for this purpose. The sky emission lines were also used to quantify the quality of the wavelength calibration, which had an average accuracy of 0.03~\AA, or 1.5~km\,s$^{-1}$.
The final step of the data reduction was the combination of all individual exposures obtained for a pointing.
The end products of the data reduction were three data cubes, one for each field. Each cube has a dimension of approximately $300\times300\times3680$ pixels.
In Fig.~\ref{fig:rgbfig}, we show the broad- and narrow-band images obtained from the reduced data cubes.

\subsection{Source extraction and photometric cleaning} 
\label{subsec:extraction}

The PampelMUSE software described in \citet{Kamann2013} was used to extract the stellar spectra from the reduced data cubes.
To run PampelMUSE, an input catalogue of source positions and an estimate of their initial magnitudes is required. 
Since the MUSE data are only moderately crowded, the input catalogues were directly created from the data cubes. We produced synthetic broad-band \textit{V-}, \textit{R-} and \textit{I-}band as well as emission line color-composite images (see Fig.~\ref{fig:rgbfig}), using the procedures of \citet{Roth2018}, and extracted photometric catalogues using the DAOPHOT code \citep{Stetson1987}. The photometric catalogues were obtained as in \citet{Gallart1996}.
They proved to be a valuable resource for our analysis, as they contain the spatial and photometric information of all the extracted sources. The details of their astrometric and photometric calibration are reported in Appendix~\ref{subsec:photometry}.

The raw source catalogues (i.e. without astrometric solution or photometric calibration applied) were passed as input to PampelMUSE to first identify the positions of stars in the data cubes as a function of wavelength. The code then aimed to retrieve the point spread function (PSF) using an analytical Moffat profile. To obtain the final spectra, a simultaneous PSF fit was performed to all sources in each layer of the data cube; unresolved components were treated by including a background in the fit that was recalculated at fixed spatial offsets.
As IC~1613 contains gaseous emission that varies on rather small spatial scales, well visible in Fig.~\ref{fig:rgbfig}, the background component was recomputed every 40 spaxels. 

PampelMUSE extracted 2293 initial spectra, examples of which are shown in Figs.~\ref{fig:spectra1}, \ref{fig:spectra2} and \ref{fig:spectra3}. For each extracted spectrum, PampelMUSE provided the associated flux uncertainty per pixel and an S/N value calculated around its central wavelength (hereafter S/N$_{\rm C}$). However, as our sources cover a wide spectral range (see Sect.~\ref{subsec:ULySS}), we calculated two further S/N indicators: one around 5500~\AA\, (S/N$_{\rm 550}$), and the other on the continuum around the CaT lines (S/N$_{\rm CaT}$). This gave us a more accurate picture of the quality of our spectra depending on the spectral type of a given star. 

The photometric catalogues were used to perform a first cleaning of the spectroscopic sample, excluding clearly non-stellar sources. We selected targets based on specific photometric parameters, including sharpness between $-0.5$ and $0.5$, and a goodness-of-fit (CHI) of less than 1. We also excluded targets with magnitudes in only one band, which were generally low S/N targets identified only in the \textit{I} band. As a result, the total number of extracted sources decreased from 2293 to 2053. 

\begin{figure*}
    \includegraphics[width=\textwidth]{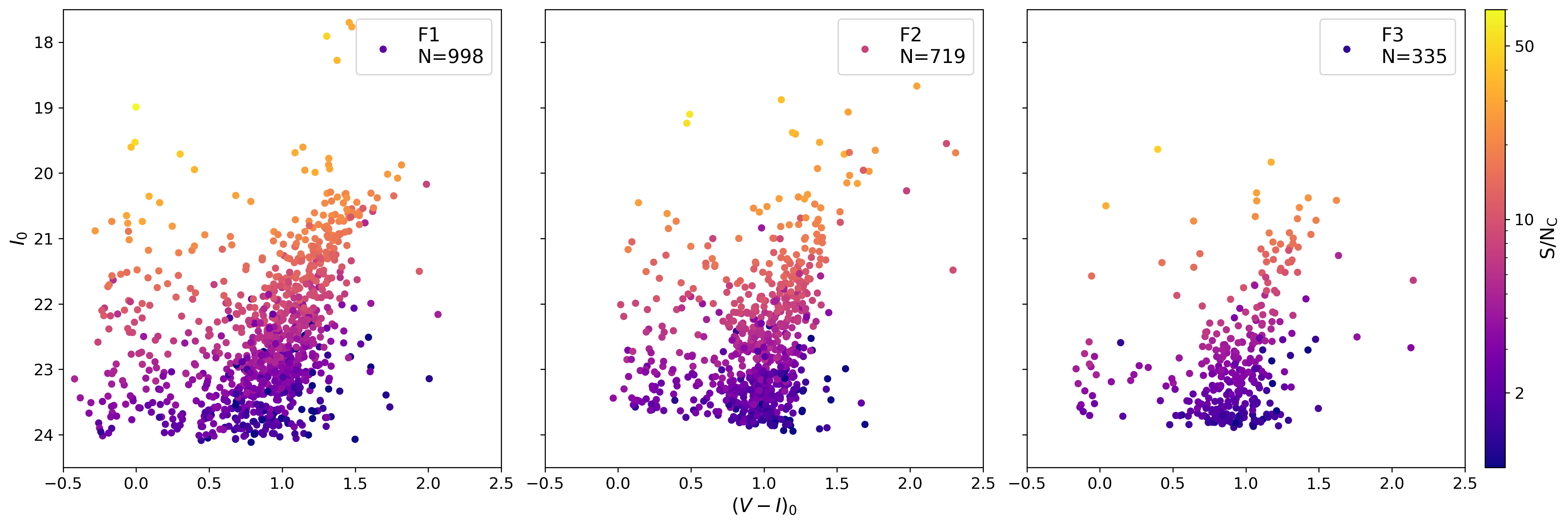}
    \caption{Extinction-reddening corrected colour-magnitude diagrams from MUSE images of the 3 fields, moving from the centre towards the outer regions from left to right; the filled circles represent the sources for which spectra have been extracted and they are colour-coded by their S/N at the central wavelength. We note that the galaxy’s distance modulus is $(m-M)_0=24.4$ \citep{Bernard2010}.}
    \label{fig:cmd_instr}
\end{figure*}

The quality of our data can be seen in Fig.~\ref{fig:cmd_instr}. In particular, we obtained respectively in F1, F2, F3 about 65, 40, 10 sources with S/N$_{550} \ge20$, and about 200, 120, 45 sources with S/N$_{\rm CaT} \ge10$.
We note that for this galaxy dust reddening and extinction along the line of sight are almost negligible, with average values of $A_V=0.067$ and $A_I=0.038$ \citep{Schlafly+Finkbeiner2011}. 
Nevertheless, we have taken this into account.

\section{Spectral types and velocity determination}
\label{sec:velocity}

We performed an analysis to determine the spectral type and line-of-sight (l.o.s.) velocity of our sources.
For this purpose, we used, respectively, the spectral fitting codes ULySS \citep{Koleva2009}\footnote{\url{http://ulyss.univ-lyon1.fr}} and \textsc{spexxy} \citep{Husser2012}.
The reason for this dual approach is motivated as follows. 

Crowded field IFS can be considered an accomplished technique for globular clusters \citep[GCs;][]{Giesers2018,Giesers2019,Latour2019,Kamann2020,Saracino2022}.
However, for galactic systems at larger distances, involving young stellar populations and ongoing star formation, the method is still at a stage of exploration. 
Unlike GCs that are dominated by old stars, complications related to hot stars, emission line stars, carbon stars, but also to unresolved star clusters and the presence of strong emission lines from H\,II regions, make it difficult at present to blindly use automated fitting tools such as \textsc{spexxy} for nearby galaxies. 
The scheme of visually assisted spectral fitting described below was first developed in a pilot study of the disk galaxy NGC\,300 at a distance of 1.88~Mpc \citep[][hereafter R18]{Roth2018}.

\subsection{Spectral classification with ULySS}

\subsubsection{General method and classification results}
\label{subsec:ULySS}

Following R18, we used ULySS to fit the observed spectra to a linear combination of 10 templates with different weights. Templates were extracted from the MIUSCAT empirical library\footnote{The MIUSCAT stellar spectral library fills in the gap between the MILES \citep{Sanchez-Blazquez2006} and the near-IR CaT library of \citet{Cenarro2001}, using the Indo-US spectral library of \citet{Valdes2004}; it also extends the wavelength coverage slightly blue-ward of MILES and red-ward of the CaT library. MIUSCAT has been recently replaced by E-MILES \citep{Vazdekis2016}.}, rather than from a model grid. 
We note that we supplemented the MIUSCAT library with 19 spectra of carbon stars from the X-Shooter library \citep{Chen2014}\footnote{\url{http://xsl.astro.unistra.fr/page\_dr1\_all.html}}. 
We also added a set of mock spectra of Be stars composed ad hoc from library B-star spectra, combining two-component Gaussians at H$_\alpha$ and H$_\beta$ wavelengths with different equivalent widths.
While these modifications did not aim to derive significant stellar parameters, they proved useful in identifying candidate stars that are not included in the MIUSCAT library for their later visual confirmation and velocity determination (see Sect.~\ref{subsec:CS+BEM}).

The linear combination fitting mode of ULySS has been employed in R18 with the experience that unresolved blends from different spectral types are flagged to prompt an inspection. However, we found that the mild crowding in IC\,1613 did not present us with such cases, unlike the more distant galaxy NGC\,300. Occasionally though, the fit returned implausible mixes of spectral types (e.g., mixing young early-types with evolved late-types) for low S/N stars. 
The apparent magnitude obtained by shifting the absolute magnitude in \textit{R-}band of the stars in the MIUSCAT library to the distance of IC~1613 (distance modulus $(m - M)_0 = 24.40$, from \citealp{Bernard2010}), provided a way to weed out these clearly erroneous fits when they would lead to apparent magnitudes well below the detection limit of the MUSE data. We also found that good-quality fits to cool MS stars allowed us to confidently identify Milky Way foreground stars that would otherwise have been difficult to reject from photometry alone.

The ULySS spectral analysis was carried out on all those sources having a S/N$_{\rm C}>2$ and a clean photometry (see Sect.~\ref{subsec:extraction}). Main outputs of the code were the best fitting templates, together with the associated spectral parameters ($T_{\rm eff}$, log(\textit{g}) and [Fe/H]) and l.o.s.~velocity. We verified by eye the ULySS outcomes, storing the information of the most probable template chosen among those with the largest weights. We also assigned several quality flags to each analysed object, evaluating the quality of the input spectrum (QSP), the quality of the spectral fitting (QFT), the plausibility of the output l.o.s.~velocity (PVR) and the plausibility of the spectral type classification (PCL). Flags were reported as integer numbers ranging from one to four, with the lower value meaning a useless measure, and the higher value implying an excellent fit. We refer the reader to R18 for further details.

In this work, we used ULySS exclusively for spectral classification purposes, while we left the l.o.s.~velocity determination part to the \textsc{spexxy} code. This was mainly because \textsc{spexxy} performed better than ULySS in the velocity determination task, especially in the low S/N regime (i.e., $\lesssim10$), as we verified in Appendix~\ref{apx:sanity-checks} and \ref{apx:spexxy-vs-ulyss}.

\begin{table}
    \caption{Distribution of spectral type classification.}
    \label{table:ULySS-spt}      
    \centering          
    \begin{tabular}{c c c}
        \hline\hline
        Sp.~Type & $T_{\rm eff}$~(K) & N \\ 
        \hline
        M & $2400-3900$   & 1 \\
        K & $3900-4900$   & 467 \\
        G & $4900-6000$   & 90 \\
        F & $6000-7500$   & 46 \\
        A & $7500-10000$  & 57 \\
        B & $10000-30000$ & 96 \\
        O & $>30000$      & 13 \\
        \hline
        Be & $> 10000$& 24 \\ 
        C  & $< 3500$ & 14  \\ 
        \hline
        Total & & 808 \\ 
        \hline
    \end{tabular}
    \tablefoot{From inspection of ULySS outcomes, with PCL flag between 3 and 4, as described in Sect.~\ref{subsec:ULySS}. 
    In the table, C and Be indicate respectively the identified number of carbon and B emission-line stars.
    }
\end{table}

\begin{figure}
    \centering
    \includegraphics[width=\columnwidth]{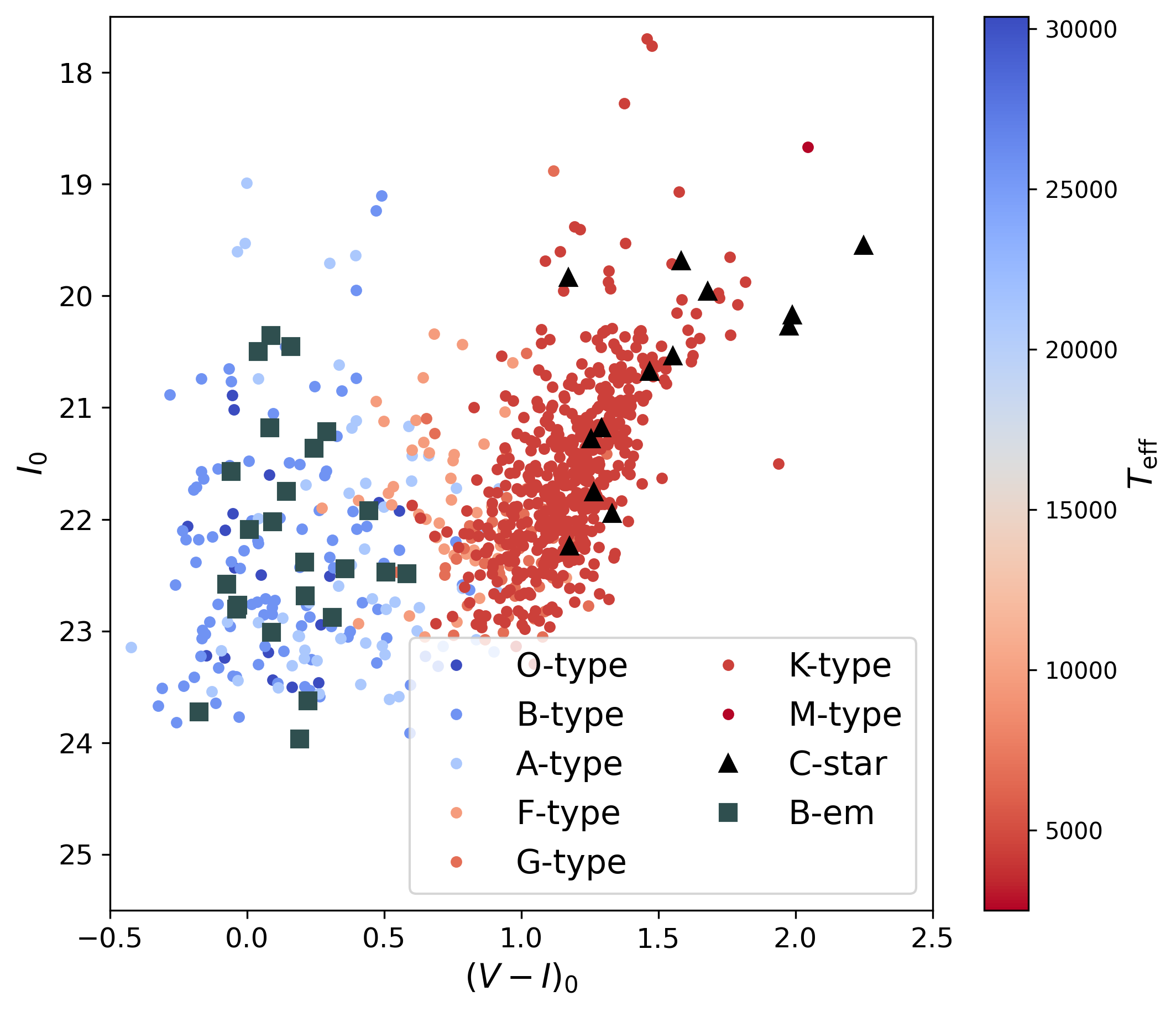} 
    \caption{Colour-magnitude diagram for data selected with PCL-flag between 3 and 4, as described in Sect.~\ref{subsec:ULySS}. Filled circles are colour-coded according to their effective temperature $T_{\rm eff}$ as derived with ULySS; black triangles indicate the identified C-stars, while the grey squares mark the Be stars.}
    \label{fig:CMD-Teff}
\end{figure}

In Table~\ref{table:ULySS-spt}, we report results from the spectral classification obtained for sources marked with a PCL flag between three and four (which implies an accuracy on the effective temperatures better than 500~K). 
The visual inspection during the spectral classification also allowed us to identify other contaminants that were not removed during the previous photometric selection (Sect.~\ref{subsec:extraction}): they were mainly background galaxies (comprising high-z emitters) and low S/N targets highly polluted by diffuse ionised gas emission lines. 
We also visually identified carbon and B emission line stars that were missed by ULySS due to the lack of adequate stellar templates, in order to study them separately (see Sect.~\ref{subsec:CS+BEM}). 
We note that we define as bona-fide B emission-line stars (noted as Be hereafter) those sources whose spectral distribution is compatible with a star having $T_{\rm eff}>10000$~K, but which show H$_{\alpha}$ and (occasionally) H$_{\beta}$ in strong emission \citep[e.g.,][]{Porter+Rivinius2003,Rivinius2013}. 
The other seven stars showed only H$_{\alpha}$ in weak emission, but due to their low S/N their classifications were too uncertain and they were therefore discarded. 

Overall, we obtained a reliable spectral classification for 808 stars\footnote{The spectroscopic $T_{\rm eff}$ for the red giant stars were found to be in fair agreement with the empirical $T_{\rm eff}:(V-I)$ relation from \citet{Alonso1999}, which is valid in the colour range $0.8 < (V-I) < 2.2$ and mostly metallicity independent.}. Looking at Table~\ref{table:ULySS-spt}, the large majority of sources are K-type giants, with an almost even representation of the other types, except for the M and O stars being a minority. In particular, among the M stars, only one was found to be a giant, while the other four were classified as MW-foreground dwarfs. We also report the presence of two other MW-contaminants classified as K\,V and G\,V stars, all confirmed in the l.o.s.~velocity determination step. We have therefore excluded these MW-foreground stars from the counts in Table~\ref{table:ULySS-spt} and the following analysis. Fig.~\ref{fig:CMD-Teff} shows the position of the classified stars on the colour-magnitude diagram, colour-coded according to their $T_{\rm eff}$, visually confirming the general goodness of the spectral classification.

\subsubsection{O and Be stars}

The number of O-type stars identified is the most uncertain, since the poor coverage of the MIUSCAT library at the highest effective temperatures (i.e. for $T_{\rm eff}>30000$~K) is a limiting factor for the spectral classification of these stars. We have identified 13 stars with a $T_{\rm eff}$ just above 30000~K. This is in general agreement with the expected number of O stars in IC~1613. Indeed, considering the galaxy's current star formation rate \citep{Skillman2014} and assuming a \citet{Kroupa2001} initial mass function, we would expect on the order of ten O stars in the surveyed area, given their average lifetime of 10~Myr and minimum mass of $16\,M_\odot$. 

\begin{figure}
	\centering
	\includegraphics[width=\columnwidth]{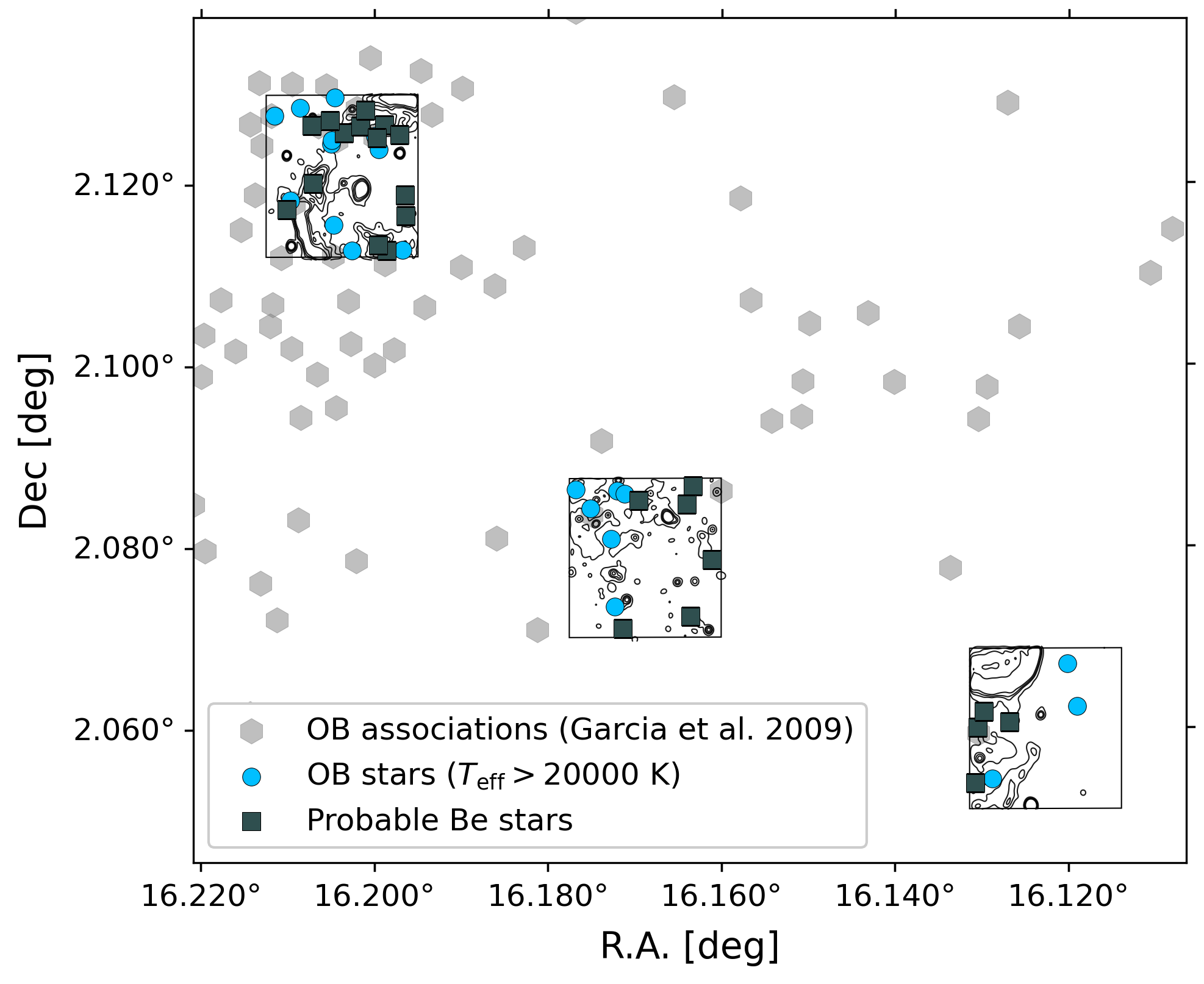}
	\caption{Spatial distribution of hot OB stars (blue filled circles) and Be stars (dark-slate grey filled squares), compared to that of the OB-associations (gray filled hexagons) reported by \citet{Garcia2009}. The black contours indicate the H$_\alpha$ emission within the MUSE pointings, marked by black squares.}
	\label{fig:BEM}
\end{figure}

On the other hand, O stars are typically not uniformly distributed in space. They tend to form and evolve in OB associations. Our observations cover some of the OB associations identified by \citet{Garcia2009}. These associations are low density groups of young stars with significant internal extinction, making the identification of O types more challenging. In Fig.~\ref{fig:BEM} we show the spatial distribution of the OB associations reported by \citet{Garcia2009}, together with that of our identified OB stars with $T_{\rm eff}>20000$~K and also the Be stars. We see that they generally tend to be found where the OB associations are. Their spatial distribution also generally follows that of the ionised shells shown in Fig.~\ref{fig:rgbfig}, to which they are probably physically associated \citep{Borissova2004,Garcia2009}. At this stage, these evidences at least confirm the goodness of our initial classification of many of them as hot stars.

We further note that the identified Be stars account for up to 18\% of the total OB sample. This is comparable to the observed Be fraction in nearby metal-poor dwarf galaxies \citep[between 15\% and 30\%,][]{Schootemeijer2022,Gull2022,Vaz2023}. Since Be stars are likely to be highly rotating massive stars, our results add to the evidence that they are common in low-metallicity environments.

We are currently preparing a follow-up study using a grid of model atmospheres obtained from the NLTE code FASTWIND \citep{Puls2005}, whose spectra will allow us to fit the more massive stars of spectral type A$\dots$O that are sparsely covered by the empirical MIUSCAT library. Nevertheless, the wavelength range of MUSE, which does not cover the gravity and temperature sensitive lines below 4800~\AA, is a limitation for the analysis of massive stars. An accurate determination of these parameters, as well as Fe and $\alpha$-elements abundances, will be possible with the future BlueMUSE instrument \citep{Richard2019arXiv}.

\subsubsection{C/M fraction}

Looking at the coolest stars in our sample, we have identified a single M-type giant and 14 carbon (C) stars. The latter are a type of asymptotic giant branch (AGB) stars, characterised by having more carbon than oxygen in their atmospheres. This happens during the thermally pulsating (TP) phase, when C-rich material from the star's interior is dredged up into the previously oxygen-rich atmosphere (a process known as the third dredge-up). A TP-AGB star is initially a O-rich M-type giant. Since it is easier to transition from an O-rich to a C-rich star when its metallicity is low, the C/M fraction is usually an indirect [Fe/H] indicator of intermediate-age stars in a galaxy. For IC~1613, literature works report an average value of C/M~$\sim0.6$ \citep[see e.g.][based on wide-area \textit{JHK}-photometric surveys]{Albert2000,Chun2015,Sibbons2015,Ren2022}.

We can qualitatively determine the C/M in our spectroscopic sample.
The only M-type giant was clearly an O-rich AGB star. We also examined the K-type stars above the tip of the RGB and selected a sample with $T_{\rm eff}$ within 500~K of an early M-type star.
Cross-correlating with the literature photometric catalogues \citep{Albert2000,Chun2015,Sibbons2015}, we found 10 common targets classified as M stars (after excluding the possible red supergiants identified by \citealp{Ren2022}). 
As for our C stars, three have already been identified in the literature \citep[see again][]{Albert2000,Chun2015,Sibbons2015}, while five have a more uncertain spectral classification due to their faint magnitudes ($I>21$), well below the tip of the RGB. It should be noted that TP-AGBs are long-period variables (with periods of a few tens to hundreds of days), which can exhibit brightness excursions of several magnitudes \citep[see e.g.,][]{Menzies2015}. Alternatively, faint C stars can be binary systems in which the primary star has received C-enriched material from the secondary star, which was previously an AGB (also known as extrinsic C stars). For IC~1613, the expected fraction of extrinsic C stars is rather low ($\sim10\%$, \citealp[e.g.,][]{Hamren2016}). 
Considering between nine and 14 C stars, the C/M ratio then ranges between 0.9 and 1.4, which is larger than the average value reported in the literature. However, taking into account the Poisson error associated with the small number statistics of our sample, we would remain in agreement with the literature. 

\subsection{Velocity determination}

\subsubsection{Velocity determination with SPEXXY}
\label{subsec:spexxy}

\begin{figure*}
	\centering
	\includegraphics[width=\textwidth]{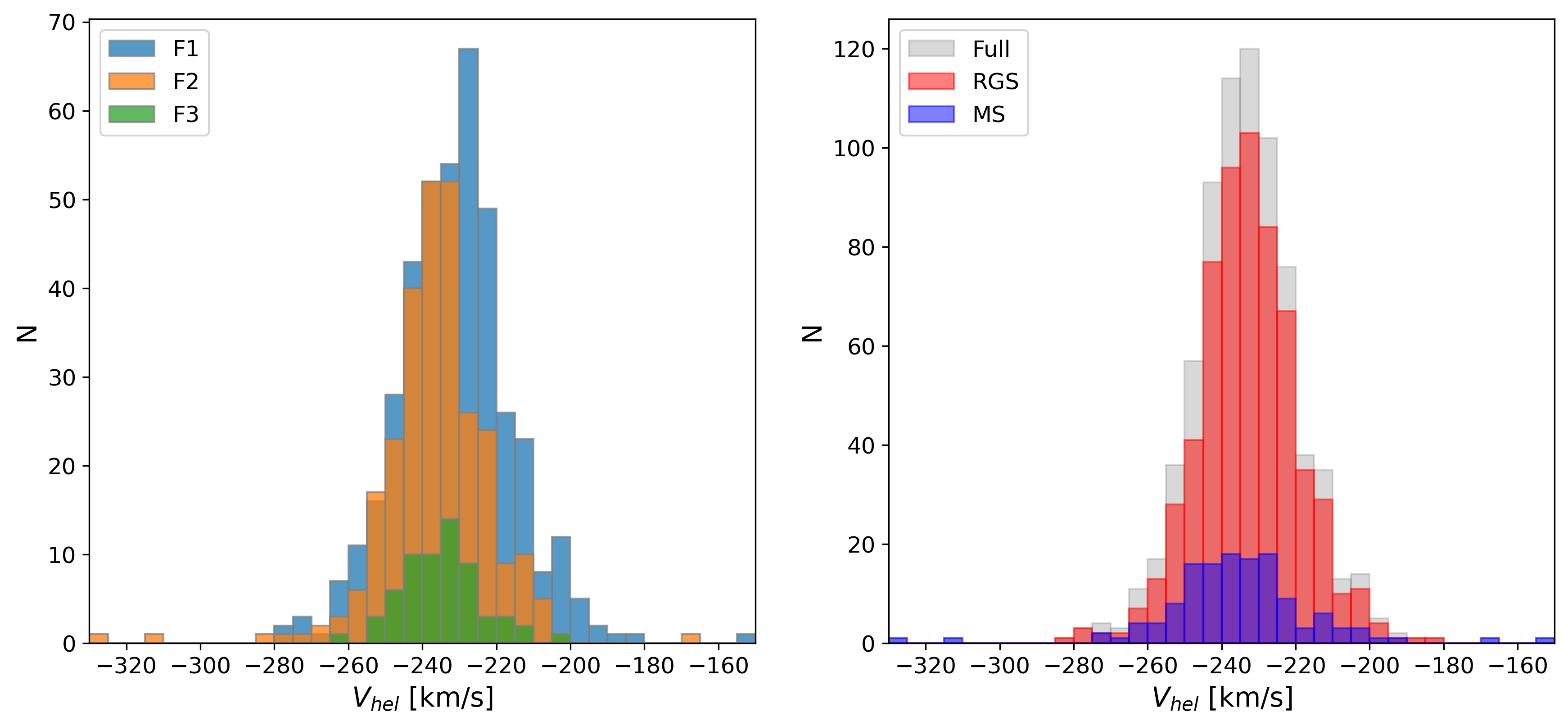}
	\caption{Histogram of the l.o.s.~velocity measurements divided by pointing (\textit{left} panel) and stellar type (\textit{right} panel).}
	\label{fig:kin_hist}
\end{figure*}

The determination of l.o.s.~velocities was carried out using the \textsc{spexxy}\footnote{\url{https://github.com/thusser/spexxy}} code \citep[v.~2.5,][]{Husser2012}. 
The routine performs a full spectral fitting to the observed spectra using interpolated spectral templates generated from the PHOENIX library of high-resolution synthetic spectra\footnote{\url{http://phoenix.astro.physik.uni-goettingen.de/}} \citep{Husser2013}. The library covers a wide wavelength range, going from 500~\AA\, to 5.5~$\mu$m, and stellar parameters: $2300<T_{\rm eff}\,(\rm K) <15000$, $0.0<{\rm log}(g)\,(\rm dex) <+6.0$, $-4.0<{\rm [Fe/H]\,(dex)}<+1.0$ and $-0.2<{\rm [\alpha/M]\,(dex)}<+1.2$. However, the PHOENIX models do not treat radiative transfer in the atmospheres of massive stars affected by strong stellar winds, so the library is of little use in the recovery of stellar parameters for $T_{\rm eff} > 10000$~K. 
We refer to Appendix~\ref{apx:sanity-checks} for a detailed explanation of the spectral fitting steps and to \ref{apx:spexxy-vs-ulyss} for a comparison of its performance with the ULySS results presented above.

We ran \textsc{spexxy} on all targets having S/N~$>2$ (in at least one of the S/N indicators) and clean photometry (see Sect.~\ref{subsec:extraction}). We also excluded all those targets that in the spectral classification step (see Sect.~\ref{subsec:ULySS}) were identified as contaminated by ionised gas, background galaxies, and Be stars (which we analysed separately). These steps reduced our sample to 1745 input sources.
We were thus able to recover their l.o.s.~velocity, along with estimates of their spectral parameters and associated errors. We note that only for $2\%$ of stars \textsc{spexxy} failed to perform a successful fitting, mostly due to their low S/N. 

The goodness of the recovered parameters was quantified in a series of tests, whose details are given in Appendix~\ref{apx:sanity-checks}. Here we report the essentials. We verified by performing several tests on mock spectra that the l.o.s.~velocity errors are well estimated down to S/N$_{\rm CaT}$~$\sim3.5$ for red giant stars (S/N$_{\rm 550}$~$>5$ for MS stars), and the velocity values are recovered without significant offsets. On the other hand, the recovery of the spectral parameters was more limited, with $T_{\rm eff}$ in particular being generally well constrained for red giants, but underestimated for the hottest MS stars due to the grid limitation of our spectral models. We emphasise again that our aim was to use \textsc{spexxy} mainly for the determination of l.o.s.~velocities. 

Since our main goal is to analyse not only the global kinematic properties of IC~1613, but also how they change as a function of stellar types, we decided to keep the sample obtained during the spectral classification step. From this sample we further removed those targets with a velocity error $\delta_{\rm v}>25$~km\,s$^{-1}$, which meant excluding $10\%$ of sources with the least reliable velocity values (mostly low S/N MS stars). 
We also excluded the six targets previously classified as MW contaminants, confirming that they are foreground sources based on their l.o.s.~velocities, which were consistent with 0~km\,s$^{-1}$. Performing a cross-correlation with the third data release of the \textit{Gaia} catalogue \citep{GaiaCollaboration2022}, we further confirmed their foreground nature based on their proper motion values, significantly different from 0~km\,s$^{-1}$ in this case.
With these conditions applied, our sample reduced to 727 sources to which we subsequently added the 24 Be stars whose velocity determination we describe in the next section. 
The distribution of l.o.s.~velocity measurements for our sample is shown in Fig.~\ref{fig:kin_hist}, with histograms divided by pointing and stellar type.
We note that the median velocity error was of $\delta_{\rm v}\sim6.5$ km\,s$^{-1}$, while the median S/N$_{\rm C}$ resulted around 10.

\subsubsection{Velocity determination for carbon and Be stars}
\label{subsec:CS+BEM}

The l.o.s.~velocities of the C stars identified by spectral classification (see Sect.~\ref{subsec:ULySS}) were calculated using \textsc{spexxy} as for the main sample. However, the code was not always able to match the complex spectral features of these stars. Therefore, we double-checked the \textsc{spexxy} velocity determination with ULySS using stellar templates from the X-Shooter library of carbon stars \citep{Gonneau2016}. We found a general agreement within the errors, confirming the goodness of the \textsc{spexxy} velocities. The median velocity error of the 14 C stars analysed (6 in both F1 and F2, and 2 in F3) was $\delta_v\sim4$~km\,s$^{-1}$ for a median S/N$_{\rm CaT}\sim15$.

On the other hand, the velocity determination for the Be stars was done separately, since \textsc{spexxy} failed in this task because such stars are not represented by the PHOENIX model atmospheres. We thus performed a cross-correlation with custom-made template spectra using the function \textit{correlation} of the \texttt{python} package \texttt{specutils}, part of the Astropy project \citep{Astropy}. 
Templates were generated as ad-hoc proxies using the spectra of two B-type stars (classified as B9III and B3III) from the MIUSCAT library \citep{Vazdekis2012}, to which we added two Gaussian profiles reproducing the strength and width of the H$_\alpha$ and H$_\beta$ emission lines as detected in our sample.
Since the spectra of our Be-star candidates in general did not show any split line profiles (except for two objects), the two-component Gaussian approximation seemed to be a good enough approach for the purpose of l.o.s.~velocity measurements.
We cross-correlated the observed spectra with the templates and assigned them the average of the derived velocities. The errors were instead assigned by a Monte Carlo process. For each template we generated mock spectra at different S/N (calculated around 5500~\AA) values $[2.5,5,7.5,10,15,25,40,60]$. For each S/N, we cross-correlated 250 mock spectra with the corresponding noise-free template and considered the median absolute deviation of the velocity distribution as the velocity error representative of that bin. The error distribution as a function of S/N was fitted by an exponential profile for each template and we used the mean profile to assign velocity errors to the observed spectra according to their S/N$_{550}$. 
For the 24 Be stars analysed (14, 6 and 4 in fields F1, F2 and F3, respectively), we recovered a median $\delta_v\sim7$~km\,s$^{-1}$ for a median S/N$_{550}\sim20$. 

We note that if we perform the kinematic analysis described in the next section on both the Be and MS star samples, we find no significant differences between them in terms of systemic velocity and velocity dispersion. Therefore, we included them in the main sample when we performed the kinematic analysis. 

\section{Internal kinematics}
\label{sec:kinematics}

\begin{table*}
	\caption{Parameters and evidences resulting from the Bayesian kinematic analysis of the main sample.}
	\label{table:multinest}
	\centering          
	\begin{tabular}{l c c c l}    
			\hline\hline
			Model & $\bar{v}_{\rm sys}$ & $\sigma_{\rm v}$ & \textit{k} & Bayes factor \\ 
			  & (km\,s$^{-1}$)   &  (km\,s$^{-1}$)   &   (km\,s$^{-1}$\,arcmin$^{-1}$) & ln($B_{\rm rot,disp}$)\\
			\hline     
                Linear rotation (HI) & $-230.9\pm0.7$ & $11.2\pm0.4$ & $1.1\pm0.3$ & 4.5 \\
                Linear rotation (optical) & $-232.2\pm0.6$ & $11.2\pm0.4$ & $1.2\pm0.3$ & 3.5 \\
                Dispersion only & $-233.3\pm0.5$ & $11.3\pm0.4$ & & \\
                \hline                
	\end{tabular}
        \tablefoot{The number of initial targets was N$_{\rm in}=751$, while that of the probable members with $P_{\rm M_j}>0.95$ was N$_{\rm P}=746$. The reported values of the kinematic parameters represent the median of the corresponding marginalised posterior distributions, with 1-$\sigma$ errors set as the confidence intervals around the central value enclosing $68 \%$ of each distributions. The values in the row labelled {\it (HI)} or {\it (optical)} were determined assuming the central coordinates and orientation of the kinematic major axis of the HI component or the stellar component, respectively.}
\end{table*}

\begin{figure}
	\includegraphics[width=\columnwidth]{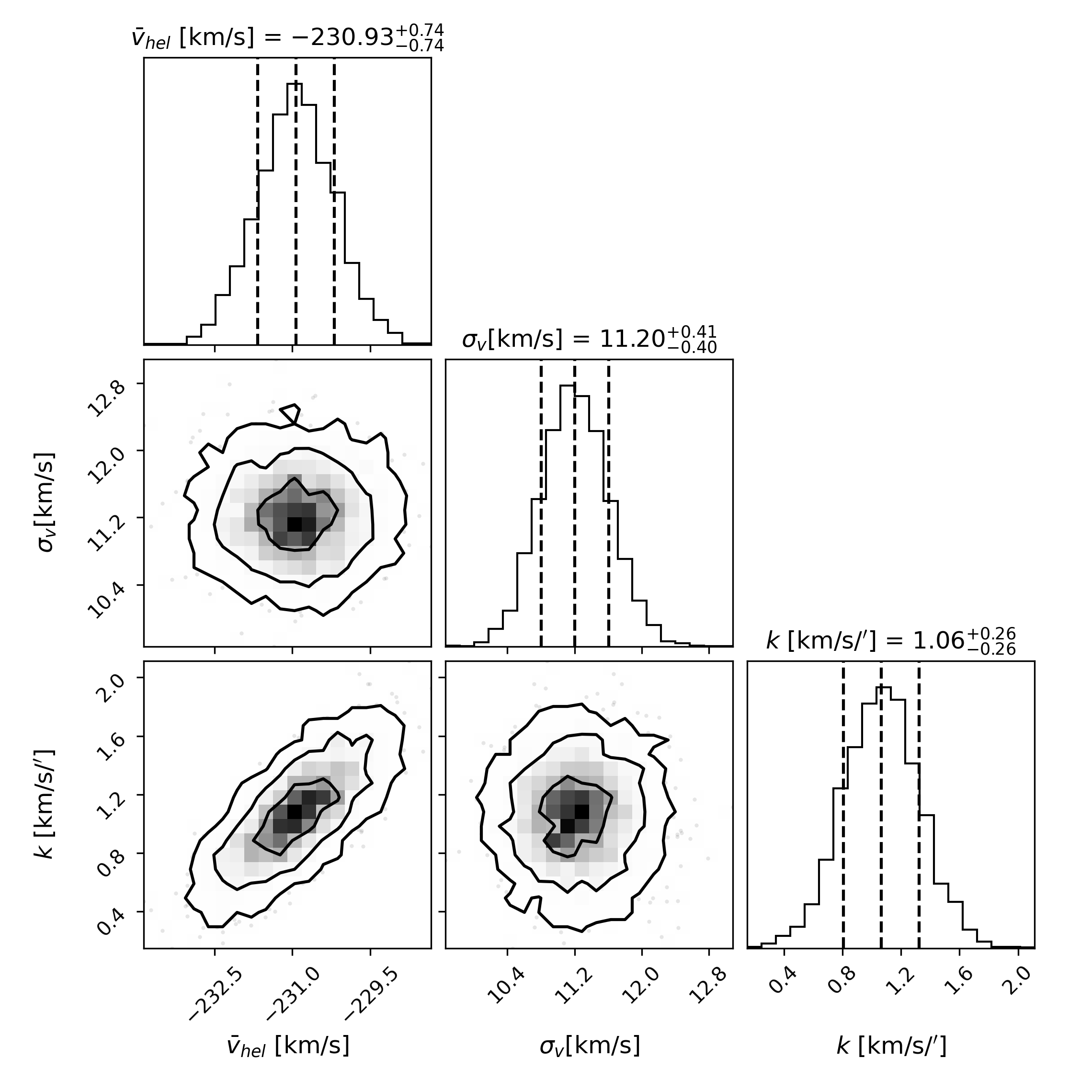}
	\caption{MultiNest 2D and marginalised posterior probability distributions for the systemic velocity, velocity dispersion, and velocity gradient assuming a linear rotational model with the kinematic P.~A.~aligned with that of the HI kinematic major axis. Dashed lines in the histograms indicate the 16th, 50th and 84th percentiles. Contours are shown at 1-, 2-, and 3-$\sigma$ level.}
	\label{fig:kin_marginal}
\end{figure}
We investigate the basic internal kinematic properties of the stellar component of IC~1613, such as its systemic l.o.s.~velocity, its velocity dispersion and the possible presence of velocity gradients, indicating rotation of the stellar component\footnote{On the angular scales covered by the MUSE sample, we find the perspective rotation due to the relative motion between the galaxy and the Sun to play a negligible role with $\Delta v$ up to $-0.2\pm0.2$~km\,s$^{-1}$ (following Eq.~6 of \citealp{VanDeVen2006} and assuming the systemic proper motion from \citealp{Battaglia2022}). Velocity gradients induced by tidal disturbances from the MW or M31 are also highly unlikely.}. 

To proceed with the kinematic analysis, we first need to identify the possible contaminants left in our sample. We had already removed many of them during the spectral classification process, with the rest expected to be faint foreground stars. We used then the method outlined in \citet{Taibi2020}, which allows a Bayesian kinematic analysis while assigning membership probabilities $P_{\rm M_j}$, assumed to be Gaussian, to the individual targets. This approach is based on the expectation maximisation technique presented in \citet{Walker2009}. 

The $P_{\rm M_j}$ values depend on the l.o.s.~velocities of the individual targets, but also on the prior information based on their radial distances from the galaxy's centre and on the expected l.o.s.~velocity distribution of the possible contaminants. 
The spatial prior takes into account that the probability of membership is higher, the closer to the galaxy's centre. In this case, we simply required that our targets should follow a monotonically decreasing radial density profile applying an isotonic regression model, with no assumptions about its functional form\footnote{\url{https://scikit-learn.org/stable/modules/isotonic.html}}. 

We carry out a Bayesian analysis to explore and compare different kinematic models, considering one where the internal kinematics of IC~1613's stellar component can be purely described by random motions, and another model which also contains a rotational term. For the rotational term we used the form: $v_{\rm rot}(R_j)\,{\rm cos}(\theta - \theta_j)$, where $R_j$ is the angular distance from the galaxy's centre (i.e. $R_j=\sqrt{\xi_j^2 + \eta_j^2}$, where ($\xi_j,\eta_j$) are the tangent plane's coordinates), $\theta_j$ is the position angle (P.~A.) of the j-target, and $\theta$ is the P.~A.~of the kinematic major axis (both P.~A.~measured from North to East). The rotational velocity term is assumed to linearly increase with radius, that is $v_{\rm rot}(R_j)=k\,R_j$.

We note that, unless otherwise stated, for the stellar component we quote the observed rotational velocity $v_{\rm rot}$ and not its intrinsic value $v_{\rm rot}^{\rm intrinsic}$, as the latter requires knowledge of the inclination angle \textit{i} of the angular momentum vector with respect to the observer, that is $v_{\rm rot}=v_{\rm rot}^{\rm intrinsic}{\rm sin}(i)$. The stellar component of LG dwarf galaxies is not in a thin disc, therefore its inclination cannot be determined without knowledge of the intrinsic 3D shape. In those systems that contain a clearly rotating HI component, as IC~1613, one can at least assume the inclination determined from the neutral gas component, under the assumptions that stars and gas share the same angular momentum vector. This is what we will do when quoting values for $v_{\rm rot}^{\rm intrinsic}$, or when deriving estimates for the circular velocity, using preferentially the value quoted in Table~\ref{table:1}. 
In Sect.~\ref{sec:incl+mdyn}, we discuss the possibility that IC~1613 may be seen close to face-on \citep{Read2016}, and suggest a new value based on the combined kinematic properties of the different stellar tracers we analysed.

\subsection{Kinematic properties of the main sample: detection of a rotation signal}
\label{subsec:kin-full}

For the kinematic analysis of IC~1613, the free parameters to be fitted were the systemic l.o.s.~velocity $\bar{v}_{\rm sys}$, the l.o.s.~velocity dispersion $\sigma_{\rm v}$, and the linear gradient \textit{k} for the rotation model. In the latter case, the particular spatial distribution of the MUSE pointings and the limited area they cover prevent us from finding $\theta$. Therefore, we fixed the central coordinates and $\theta$ of the kinematic field to be either that of the stellar component or that of the HI gas (see values in Table~\ref{table:1}). 

We assumed the following prior ranges: $-50 < (\bar{v}_{\rm sys} - v_{\rm glx}) [ {\rm km\,s^{-1}}] < 50$, where $v_{\rm glx}$ is the mean value of the input velocity distribution, $0 < \sigma_{\rm v}  [{\rm km\,s^{-1}}] < 50$, $-10 < k [ {\rm km\,s^{-1}\,arcmin^{-1}}]< 10$. 
The analysis also returned the model evidence \textit{Z} which resulted useful for comparing the statistical significance of one model against another through the use of the Bayes factor ${\rm ln}(B_{1,2})={\rm ln}(Z_1/Z_2)$. 
We obtained the kinematic parameters and the membership probabilities using the MultiNest code \citep{Feroz2009,Buchner2014}, a multi-modal nested sampling algorithm.  

We used the Besan\c{c}on model of Galactic foreground stars \citep{Robin2003} as a prior over the expected l.o.s.~velocity distribution of the contaminants. We generated a catalogue along IC~1613's direction, over an area up to its half-light radius, covering the range of colours and magnitudes of our targets. 
This resulted in a velocity distribution that was well sampled and representative of the spatial area around the MUSE pointings. We approximated it as the sum of two Gaussian profiles (with means $\bar{v}_{\rm Bes, 1} = -12$ km\,s$^{-1}$, $\bar{v}_{\rm Bes, 2} = -66$ km\,s$^{-1}$, and standard deviations $\sigma_{\rm Bes, 1} = 34$ km\,s$^{-1}$, $\sigma_{\rm Bes, 2} = 114$ km\,s$^{-1}$, with an amplitude ratio of $k_1/k_2 \sim 2$). According to this model, considering the small area covered by the MUSE pointings and the spanned range of magnitudes and colours, we expect to find, in total, around five foreground contaminants in the velocity range between $-400$~km\,s$^{-1}$ and 200~km\,s$^{-1}$. If we restrict to the velocity range covered by the input targets (i.e. between $-330$~km\,s$^{-1}$ and $-140$~km\,s$^{-1}$, see Fig.~\ref{fig:kin_hist}), we would expect only one contaminant. 
We recall that already during the spectral classification step (see Sect.~\ref{subsec:ULySS}) we found and removed six likely contaminants, all being cold MS stars with velocities around $\bar{v}_{\rm Bes, 1}$.

Results of the probability-weighted analysis are reported in Table~\ref{table:multinest}. Of the initial 751 targets, 746 have a $P_{\rm M_i}>0.95$. We found that rotation is strongly favoured over a dispersion-only model, with a higher evidence for the case that assumes the central coordinates and orientation of the kinematic major axis from the HI. In this case, the l.o.s.~systemic velocity of the model including a rotation term is $-230.9\pm0.7$~km\,s$^{-1}$, the velocity dispersion $11.2\pm0.4$~km\,s$^{-1}$, and the velocity gradient was $1.1\pm0.3$~km\,s$^{-1}$\,arcmin$^{-1}$ (see Fig.~\ref{fig:kin_marginal}). As these values have the highest evidence, we adopt these as our reference values in the following analysis.
 
The systemic velocity and velocity dispersion values we found are within 1-$\sigma$ from those previously published by \citet[][i.e. $-231.6\pm1.2$~km\,s$^{-1}$ and $10.8^{+1.0}_{-0.9}$~km\,s$^{-1}$, respectively]{Kirby2014}. 
In contrast to this previous work, we detect a clear rotation signal. In particular, \citet{Wheeler2017}, analysing data from \citet{Kirby2014}, reported a rotation amplitude for IC~1613 that was largely unconstrained and consistent with a null value. This is not surprising as these data are spatially distributed along the kinematic minor axis (i.e. roughly perpendicular to the direction of our MUSE dataset). Therefore, this is the first time that stellar rotation is detected with high significance in IC~1613.

We note that the five low probability targets could still belong to IC~1613 based on their spectral type. Only one has $P_{\rm M_i}\approx0$, but we checked that the velocity determination in this case was affected by the presence of residual emission lines from the surrounding ionised gas. By avoiding these lines, we recalculated its velocity, which was compatible with $\bar{v}_{\rm sys}$, although with too large an error ($\delta_v\sim30$~km\,s$^{-1}$). The other stars have $0.3<P_{\rm M_i}<0.95$. Three of these are low S/N young stars with large velocity errors. Interestingly, the remaining star has $P_{\rm M_i}\sim0.8$ and $v_i = -184\pm6$~km\,s$^{-1}$, which puts it more than 3$\sigma$ away from the expectation of the best fitting rotation model. However, it is found in F1 and has photometric properties and metallicity (see Sect.~\ref{sec:metallicity}) in common with the RGB stars of the galaxy. Given the low probability of being a MW contaminant, this star could be a possible binary, but we cannot exclude that it could belong in projection to an extended hot stellar halo around IC~1613. Indeed, there is growing evidence that MW satellites have extended stellar halos \citep[e.g.][]{Chiti2021,Longeard2023,Sestito2023a,Sestito2023b,Jensen2024}. Additionally, \citet{Pucha2019} have shown that the evolved stellar population of IC~1613 extends to $4\times R_e$.

We continue our analysis by selecting sources with a $P_{\rm M_i}>0.95$, creating a high fidelity sample without the need to recalculate individual membership probabilities. This allows us to explore the kinematic properties of different sub-samples and make a detailed comparison with the HI kinematic field. 

\subsection{Kinematic properties of each pointing}
\label{subsec:kin-fields}

\begin{table}
    \caption{Results from the Bayesian kinematic analysis performed for each pointing and stellar type applying a dispersion-only model.}
    \label{table:multi2}
    \centering          
    \begin{tabular}{c c c c c}    
	\hline\hline
	\multicolumn{2}{c}{Sample} & N$_{\rm P}$ &$\bar{v}_{\rm Field}$ & $\sigma_{\rm v}$ \\ 
	    &  &  & (km\,s$^{-1}$) & (km\,s$^{-1}$) \\
	\hline     
             & F1 & 410 & $-231.3\pm0.7$ & $12.1^{+0.6}_{-0.5}$ \\
	Main & F2 & 274 & $-236.2\pm0.7$ & $9.0\pm0.6$ \\
             & F3 & 62  & $-234.5\pm1.3$ & $8.2^{+1.2}_{-1.0}$ \\  
             & Total & 746 &        &  \\
             &    &     &                & \\
            & F1 & 331 & $-230.6\pm0.7$ & $12.2\pm0.6$ \\
	RGS & F2 & 230 & $-236.1\pm0.8$ & $ 9.3\pm0.6$ \\
            & F3 & 54  & $-234.4\pm1.4$ & $8.6^{+1.2}_{-1.1}$ \\ 
            & Total & 615 & -- &  $11.2\pm0.4$  \\
            &    &    &    & \\
  		& F1 & 79 & $-234.9^{+1.6}_{-1.6}$ & $11.8^{+1.5}_{-1.3}$ \\
	MS   & F2 & 44 & $-237.3\pm1.6$ & $6.2^{+2.0}_{-1.6}$ \\
		& F3 & 8  & $-236.6^{+4.2}_{-2.8}$ & $5.2^{+5.2}_{-3.2}$ \\  
            & Total & 131 & -- & $10.2^{+1.1}_{-1.0}$ \\
            &    &    &    & \\
  		& F1 & 84 & $-230.9\pm1.3$ & $11.3^{+1.0}_{-0.9}$ \\
	RGB-MR   & F2 & 44 & $-237.0\pm1.4$ & $8.1^{+1.3}_{-1.1}$ \\
		& F3 & 10 & $-237.5^{+1.7}_{-1.5}$ & $3.5^{+1.8}_{-1.4}$ \\  
            & Total & 138 & --  & $10.3\pm0.7$ \\
            &    &    &    & \\
  		& F1 & 70 & $-229.1^{+1.7 }_{-1.6}$ & $13.0^{+1.3}_{-1.2}$ \\
	RGB-MP   & F2 & 50 & $-236.8^{+1.5 }_{-1.4}$ & $9.6^{+1.3}_{-1.1}$ \\
		& F3 & 17 & $-231.8^{+2.9}_{-3.1}$ & $11.6^{+2.7}_{-2.2}$ \\  
            & Total & 137 &  --  & $12.0^{+0.9}_{-0.8}$ \\                        
	\hline
    \end{tabular}
    \tablefoot{The results refer to the main sample, as well as to the subsamples of RGS and MS stars defined in Sect.~\ref{subsec:kin-spType}, and to the MP and MR subsamples of RGB stars with metallicity measurements defined in Sect.~\ref{sec:AVR}. The values of the kinematic parameters and their associated uncertainties were calculated as reported in Table~\ref{table:multinest}.}
\end{table}

\begin{figure*}
	\centering
	\includegraphics[width=\textwidth]{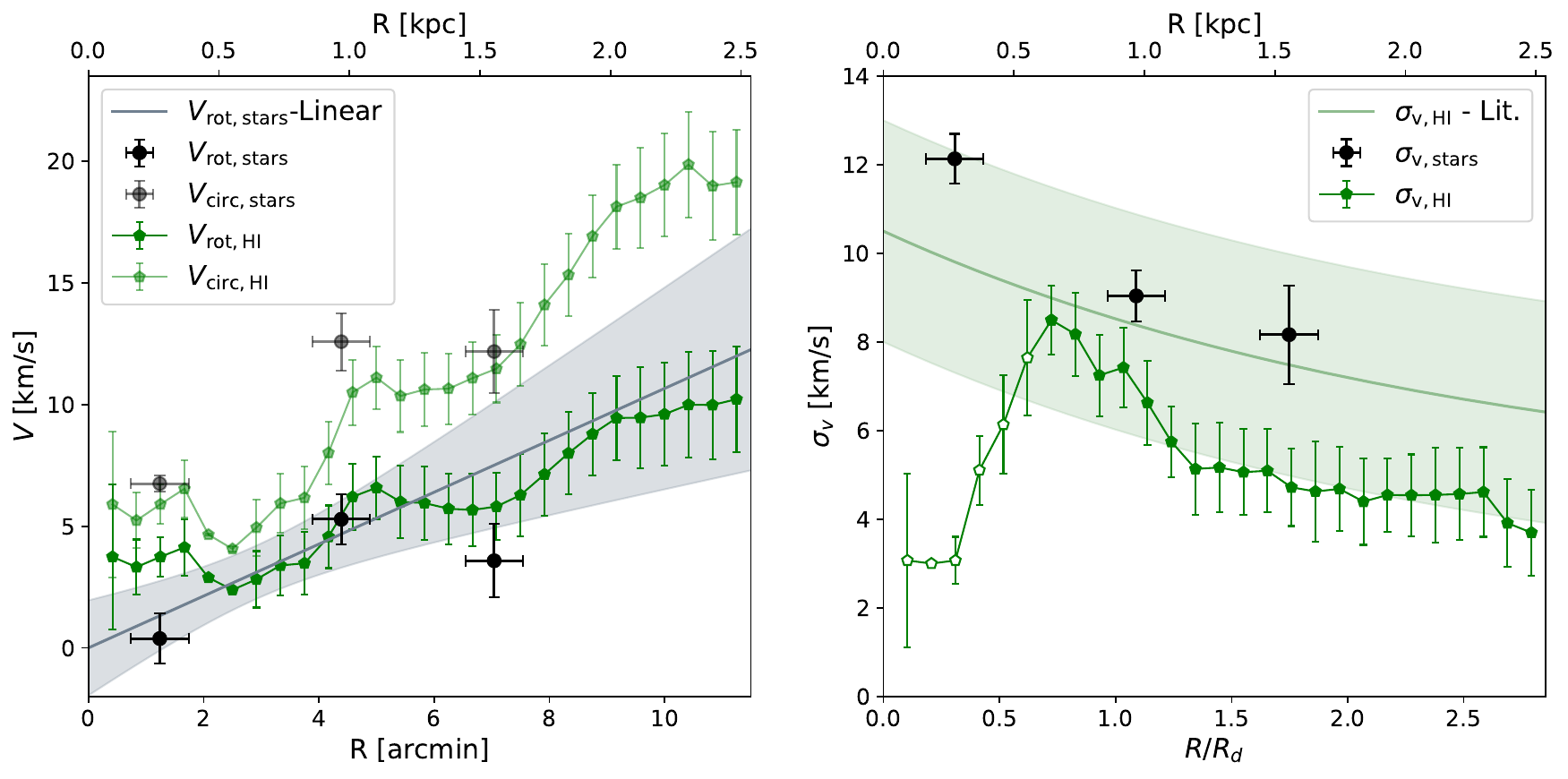}\\
        \includegraphics[width=\textwidth]{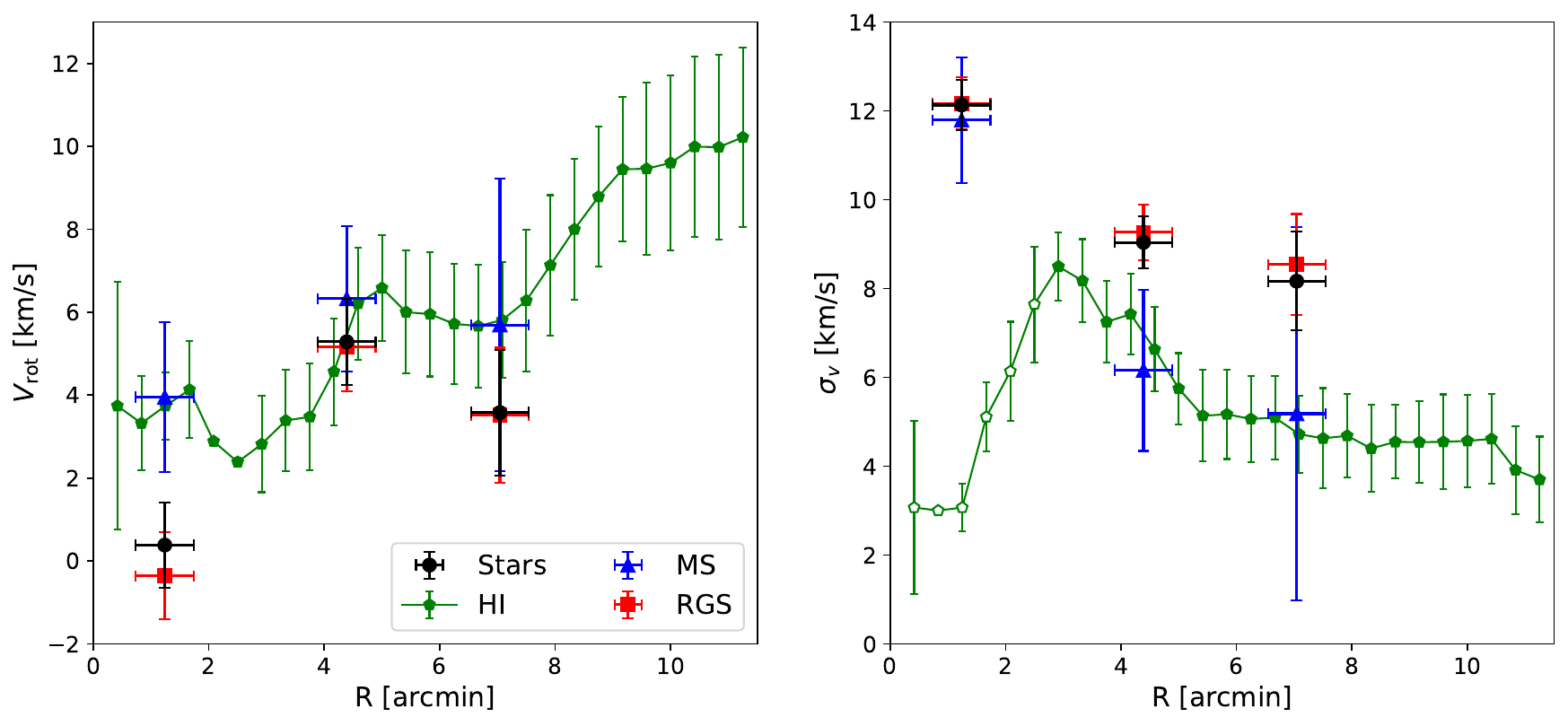}
        \caption{Average rotational velocity (\textit{left}) and velocity dispersion (\textit{right}) values for the full stellar sample (\textit{top}) and dividing by spectral type (\textit{bottom}) projected along the HI kinematic semi-major axis. Black circles represent the values obtained for the probable members per pointing, while coloured symbols indicate values obtained from the sub-samples of red giant (red squares) and main sequence stars (blue triangles); also shown are the rotational velocity and dispersion values for the HI (green pentagons; \citealp{Read2016}). The circular velocity values for both stars and the HI are also shown (light black circles and light green pentagons, respectively), correcting for the asymmetric drift and assuming an inclination angle of $i_{\rm HI}=39.4^\circ$. 
The grey line represents the rotation curve from the linear model adopted in Sect.~\ref{sec:kinematics}, with the grey bands indicating the 95\% confidence interval. 
The light-green line is an exponential fit to the dispersion profiles of a set of dwarf galaxies with high-resolution HI data having similar mass as IC~1613 and radii normalised to the scale radius $R_{\rm d}$ \citep{Iorio2017,Mancera-Pina2021,Mancera-Pina2024}; the light-green bands indicates a typical uncertainty of 2.5~km\,s$^{-1}$.
For the stellar component, the error bars in the radial direction indicate that the rotational velocities are averages within the extent of each MUSE pointing.}
	\label{fig:kin_rotcurve}
\end{figure*}

Here we determine the kinematic properties of IC~1613's stellar component as a function of radius by analysing the l.o.s.~velocities of probable members within each MUSE pointing independently using a dispersion-only model. Results are reported in Table~\ref{table:multi2} and shown in Fig.~\ref{fig:kin_rotcurve}, where we also compare them to the kinematic properties of the HI component, derived in \citet{Read2016} from the Little-THINGS survey data \citep{Hunter2012,Oh2015}. To make a direct comparison between gas and stars, we use the same central coordinates, in this case those of the HI, as there is a slight offset with the optical values (see Table~\ref{table:1})\footnote{We note that throughout the text, velocities with the sub-script {\it rot} are the observed rotational velocities, neither corrected for the inclination ${\rm sin}(i)$ nor for the asymmetric drift; those with the sub-script {\it circ} are circular velocities, thereby corrected for both effects.}. We verified that the bias introduced by the rotation in the recovery of the velocity dispersion values in each field is negligible in our case \citep[see also][]{Leaman2012}. 

It is evident that the typical velocities of stars in each field $\bar{v}_{\rm Field}$ change with respect to the systemic velocity $\bar{v}_{\rm sys}$ when moving outward from the galaxy's centre (see the left panels of Fig.~\ref{fig:kin_rotcurve}, where we define the rotational velocity as $v_{\rm rot}$ as $\bar{v}_{\rm sys}-\bar{v}_{\rm Field}$ in order to show a positive trend). F1 and F2 contain almost 90\% of the inspected stars and thus dominate the linear rotation signal recovered in the analysis of the entire sample (shown as a black solid line in figure). Indeed, the velocity difference between them is $\sim4$~km\,s$^{-1}$ which, considering that the average distance between F1 and F2 is $\sim4$~arcmin, is in agreement with the observed velocity gradient of $k\sim1$~km\,s$^{-1}$\,arcmin$^{-1}$.

From Fig.~\ref{fig:kin_rotcurve}, we see that F2 and F3 show approximately the same rotational velocity (since their $\bar{v}_{\rm F2} \approx \bar{v}_{\rm F3} \approx -235$~km\,s$^{-1}$). Due to the limited spatial sampling, it is difficult to determine whether the stellar velocity signal has reached a constant value or whether it will continue to vary radially.
We recall that the only available spectroscopic dataset that we could use to improve the spatial sampling is that published by \citet{Kirby2014} which, unluckily, is mainly distributed along the optical minor axis of the galaxy providing, as already shown by \citet{Wheeler2017}, a poor constraint to the rotation signal. 
 A similar behaviour at these radii is also seen in the HI component, where the local depressions of the rotation curve may be due to the presence of large HI holes, particularly evident around 0.5 and 1.5~kpc from the centre, as shown in Fig.~\ref{fig:fov} and also discussed in \citet{Read2016} and \citet[][]{Collins+Read2022}.
 
Inspection of the velocity dispersion of the stellar component reveals a radial decrease, with the $\sigma_v$ of the central pointing being close to the value obtained from the full sample, while for F2 and F3 we find a significant decrease to a roughly similar lower value. 
The velocity dispersion profile of the HI starts at $3-4$~km\,s$^{-1}$ near the centre of the galaxy, then rises up to $\sim9$~km\,s$^{-1}$ at $R\sim0.5~R_{\rm e}$ (i.e., around the position of F2), to finally decrease and stabilise around $\sim5$~km\,s$^{-1}$ for $R\sim R_{\rm e}$. 
The velocity dispersion of the HI is systematically lower than that obtained for the stars at the same radii, with this difference being much more marked in F1. Again, it is likely that the sudden drop in the central part of the HI velocity dispersion profile (marked with hollow pentagons in Fig.~\ref{fig:kin_rotcurve}) is due to the much lower gas density in the inner region, and thus the presence of a shell-like structure within $R\lesssim0.5$~kpc; \citealp[see again Fig.~\ref{fig:fov}, but also][]{Moiseev2012,Stilp2013}).
The HI velocity dispersion is indeed correlated to the gas surface mass density, which regulates the amount of local turbulence \citep[e.g.][]{Bacchini2020}. 

To corroborate the idea that the sudden drop in velocity dispersion for the HI is related to the HI hole, we also overplotted in Fig.~\ref{fig:kin_rotcurve} (upper-right panel) an exponential fit to the dispersion profiles of a set of dwarf galaxies with high-resolution HI data having similar mass as IC~1613 (i.e. having $v_{\rm circ}<35$~km\,s$^{-1}$), as modelled in \citet{Iorio2017} and \citet{Mancera-Pina2021,Mancera-Pina2024}. In fact, the dispersion profile is expected to increase exponentially towards the centre, and it seems that this was also the case for IC~1613 before the formation of the HI hole.

In order to estimate the galaxy's circular velocity $v_{\rm circ}$ and verify that we can directly compare the kinematics of stars and gas, it is necessary to account for the inclination and remove the random motion component that suppresses the rotation curve. We followed the formalism of \citet[][see further details in Sect.~\ref{sec:incl+mdyn}]{Read2016} to apply the so-called asymmetric drift correction, assuming the same inclination angle (see Table~\ref{table:1}) for both tracers. After applying this correction, the stellar and HI components resulted in excellent agreement, as shown in Fig.~\ref{fig:kin_rotcurve} (upper left panel). 
We discuss further in Sect.~\ref{sec:incl+mdyn} the caveats related to the inclination value and their impact on the circular velocity and dynamical mass estimations.

\subsection{Kinematic properties as a function of stellar type}
\label{subsec:kin-spType}

We now divide the main sample of probable member stars into two sub-samples, young MS stars and evolved red giant stars, to examine the kinematic properties as a function of stellar type. Relying on the analysis in Sect.~\ref{subsec:ULySS}, we perform a simple selection according to the measured $T_{\rm eff}$: red giant stars (RGS, i.e., belonging to the RGB and AGB) as those with $T_{\rm eff}\leq7500$~K, and MS stars (including the Be stars) as those with $T_{\rm eff}>7500$~K. 
We verified that the $T_{\rm eff}$ limit is in general effective in separating the two samples, comparing with the $T_{\rm eff}$ expected from the $(V-I)$ colours of our targets using the empirical calibration for red giant stars from \citet{Alonso1999}, valid in the colour range of $0.8<(V-I)<2.2$ and mostly metallicity independent.
We note that these two samples are also distinct in age. Based on stellar evolution theory, it is known that the RGS track stars of ages $>1-1.5$~Gyr. While a simple comparison with a set of isochrones (with ${\rm [Fe/H]}=[-1.3; -0.7]$~dex, \citealp[][]{Girardi2000,Bressan2012}) tells us that the stars we classify as MS are between $35<t_{\rm age}[{\rm Myr}]<560$.

Results of the kinematic analysis are reported in Table~\ref{table:multi2} and Fig.~\ref{fig:kin_rotcurve}. The outcomes for the RGS are very similar (within 1-$\sigma$) to those of the full sample in terms of rotation velocities and velocity dispersion per field, as well as the global values of the linear velocity gradient. This is expected, as this sub-sample represents a large fraction ($\sim85\%$) of the main one. As is evident from Fig.~\ref{fig:kin_rotcurve}, a model allowing for a flat rotation curve, or a more steeply rising rotation curve than the one we adopted, would have been a much better match to the kinematics of the MS stars. At all radii, the MS stars are showing a higher amplitude in their rotation with respect to the RGS (and the full sample), and in F2 and F3 also a significantly lower velocity dispersion. 
From the same figure, we can see that the MS stars follow the HI kinematic properties very closely (apart from the velocity dispersion in F1), while on the other hand, the kinematics of the older RGS, have decoupled from those of the HI. Therefore, the kinematics of the young MS stars appear to be still coupled to that of the HI component, although the large errors in the external pointings weaken this conclusion.

We have verified that the reported kinematic values are generally robust against the presence of possible biases, such as the presence of variable or binary stars, as reported in Appendix~\ref{apx:kin-bias}. Only for the MS stars in F1, we found that the high fraction of OB stars present could inflate the value of the velocity dispersion towards a higher value. This would partly explain why this value is so close, in absolute terms, to that of the RGS. Nevertheless, it is insufficient to account for the large discrepancy with the HI central velocity dispersion. The simplest explanation for this discrepancy is again related to the presence of the central HI hole, where F1 is located. Extrapolating the HI velocity dispersion from the outer parts towards the centre on the basis of the trend observed in the literature for galaxies of similar mass as IC~1613 (see Fig.~\ref{fig:kin_rotcurve}), we would expect the HI velocity dispersion at the location of the central pointing to be very similar to that measured for the RGS and MS stars. 
Therefore it is possible that the dispersion and spatial distribution of the MS stars in the central pointing reflects that of the HI gas from which they were born. Such a HI hole could have been caused through stellar feedback mechanisms like a supernova (SN) explosion. The shape and nature of the central ionised shells shown in Fig.~\ref{fig:rgbfig} are typical of a SN remnant, which seem to support this explanation. In addition, the spatial distribution of the OB stars inside and just outside F1 (Fig.~\ref{fig:BEM}) seems to suggest not only a physical association with the ionised shells, but that they may have originated from the compression of cold gas due to SN shock waves \citep{Borissova2004,Garcia2009}.

\section{Metallicity properties}
\label{sec:metallicity}

\begin{table*}
	\caption{Results from the chemical analysis of RGB stars in IC~1613.}
	\label{table:met}
	\centering          
	\begin{tabular}{l c c c c c c c c c}    
		\hline\hline
		  &  &\multicolumn{4}{c}{\citet{Starkenburg2010}} & \multicolumn{4}{c}{\citet{Carrera2013}} \\ 
  		Sample & N & ${\rm [Fe/H]}$ & $\sigma_{\rm MAD}$ & $\sigma_{\rm intrinsic}$ & $\nabla_{\rm [Fe/H]}$ & ${\rm [Fe/H]}$ & $\sigma_{\rm MAD}$ & $\sigma_{\rm intrinsic}$ & $\nabla_{\rm [Fe/H]}$ \\
            & & (dex) & (dex) & (dex) & (dex $R_{\rm e}^{-1}$) & (dex) & (dex) & (dex) & (dex $R_{\rm e}^{-1}$) \\
		\hline     
		Full & 275 & $-1.06$ & 0.29 & 0.26 & $-0.06\pm0.08$ & $-1.14$ & 0.37 & 0.34 & $-0.05\pm0.11$\\
		F1   & 150 & $-1.03$ & 0.29 & 0.26 &  & $-1.09$ & 0.36 & 0.32 & \\
		F2   &  93 & $-1.10$ & 0.31 & 0.26 &  & $-1.15$ & 0.35 & 0.36 & \\
		F3   &  27 & $-1.15$ & 0.22 & 0.21 &  & $-1.22$ & 0.30 & 0.32 & \\          
		\hline
	\end{tabular}
    \tablefoot{From left to right, columns are the considered sample, its size N, its median metallicity together with its scaled median absolute deviation, intrinsic scatter, and radial metallicity gradient. Values are given for both calibration methods applied to obtain them.}
\end{table*}

\begin{figure*}
	\centering
	\includegraphics[width=0.45\textwidth]{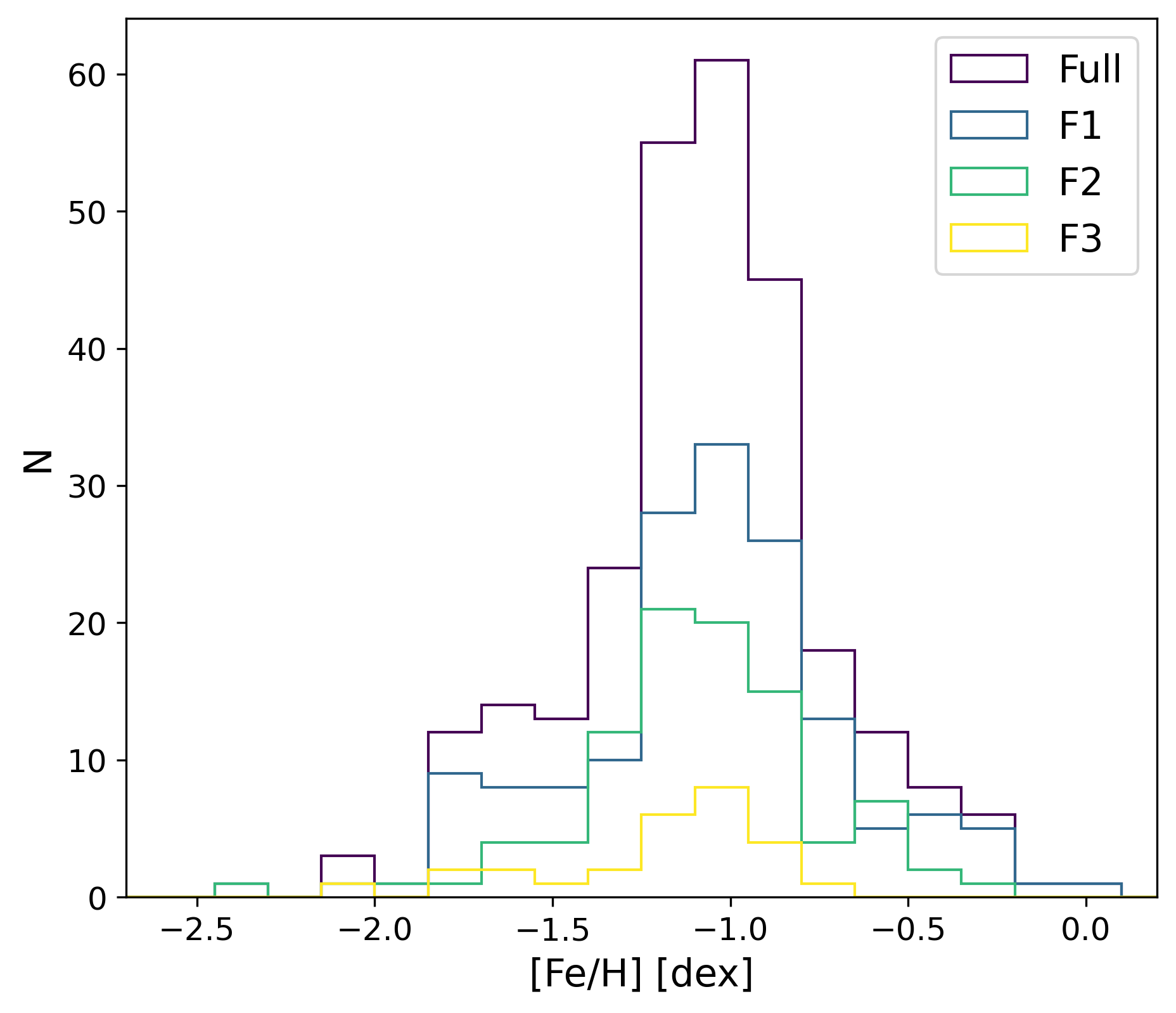}
        \includegraphics[width=0.53\textwidth]{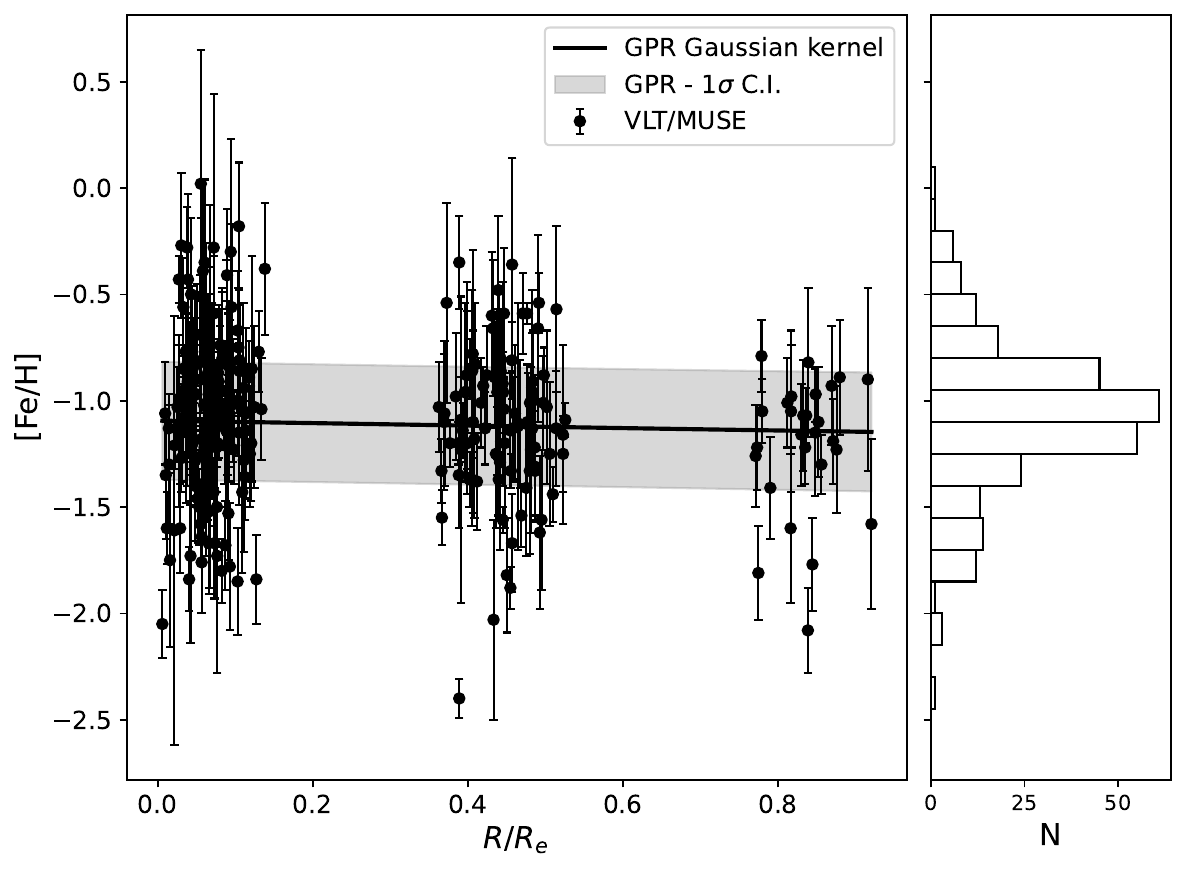}
	\caption{Metallicity distributions. \textit{Left:} histograms of the metallicity values for the full sample and for each pointing coloured in purple, blue, green and yellow, respectively. \textit{Right:} [Fe/H] values as a function of the elliptical radius scaled with $R_{\rm e}$, represented as black dots. The black solid line represents the result of a Gaussian process regression analysis using a Gaussian kernel and taking into account an intrinsic scatter; the grey band indicates the corresponding 1-$\sigma$ confidence interval. The histogram on the right side represents the metallicity distribution of the full sample.}
	\label{fig:met1}
\end{figure*}

The chemical analysis focuses only on the RGB stars in our sample, because, due to the low resolution of the MUSE data, it is difficult to make a quantitative estimate of spectral parameters other than $T_{\rm eff}$ using the spectral fitting techniques implemented in this work (see details in Appendix~\ref{apx:sanity-checks}).
Therefore, we determined metallicities ([Fe/H]) for individual RGB stars from the equivalent widths of the near-IR Ca II triplet (CaT) lines. 
We restrict our analysis to this subsample by selecting stars with $P_{\rm M_i}\geq0.95$ and $3500<T_{\rm eff}[{\rm K}]<6000$, which helps separating them from the hotter evolving giants. We verified the goodness of this selection using a set of isochrones \citep{Bressan2012} with $Z=0.0001$ ([Fe/H]$\sim -2.3$~dex) and $t_{\rm age}>1$~Gyr, which traced the lower range of metallicities and stellar ages for the RGB stars as obtained from the SFH analysis of IC~1613 \citep{Skillman2014}. We also selected all stars with S/N$_{\rm CaT}>10$ to keep the uncertainties on the equivalent widths (EWs) less than 20\%. 
This resulted in a sub-sample of 275 input sources.

We made use of the \citet{Starkenburg2010} calibration.
As shown in \citet{Kacharov2017}, this calibration can be safely applied to data of spectral resolution as low as $R\sim2600$, comparable to that of our MUSE spectra around the CaT lines.
The \citeauthor{Starkenburg2010} calibration combines the EWs of the two reddest Ca~II lines with the $(V-V_{\rm HB})$ term, where \textit{V} is the visual magnitude of the selected star and $V_{\rm HB}$ that of the horizontal branch. The EWs were obtained by fitting a Voigt profile over a 15~\AA\, window around the selected Ca~II lines, weighting each pixel value for the flux uncertainty stored in the error spectrum. The EW uncertainties were then calculated from the covariance matrix of the fitted Voigt parameters. We adopted a $V_{\rm HB}$ value of $24.91\pm0.12$ \citep{Bernard2010}. Final uncertainties for the metallicity values were obtained by error propagation of the EW uncertainties. We verified that the input magnitudes' errors do not have a significant impact on the final [Fe/H] uncertainties.
 
The uncertainties obtained in the EW measurements resulted to be about 0.5~\AA, while the average [Fe/H] error was $\sim 0.25$ dex. However, a few targets ($<5\%$) have metallicity values with errors $\delta_{\rm [Fe/H]}>0.5$~dex. We note that a small fraction of the stars in our sub-sample are likely to be AGB stars, brighter on average, for which the calibration method can be applied anyway, since it does not introduce a significant systematic bias compared to RGB stars (see discussion in \citealp{Pont2004}). 

From the analysis of the metallicity values we obtained a median ${\rm [Fe/H]} = -1.06$~dex, a median absolute deviation $\sigma_{\rm MAD} = 0.29$~dex, and an intrinsic scatter $\sigma_{\rm intrinsic} = 0.26$~dex \citep[following Eq.~8 of][]{Kirby2013}, as also reported in Table~\ref{table:met}. For comparison, we also applied the \citet{Carrera2013} calibration using the fitting windows defined by \citet{Cenarro2001}. The main difference with the \citeauthor{Starkenburg2010} calibration is that this one is empirical, whereas the former also used synthetic spectra. Nevertheless, we obtained comparable results: median ${\rm [Fe/H]} = -1.14$~dex, $\sigma_{\rm MAD} = 0.37$~dex and $\sigma_{\rm intrinsic} = 0.34$~dex (see again Table~\ref{table:met}). The recovered median metallicity is $\sim0.1$~dex lower, but the offset is within the scatter of the data, in this case $\sim0.1$~dex higher.

Our results are in good agreement with those reported by \citet{Kirby2013}: median ${\rm [Fe/H]} = -1.22$~dex and $\sigma_{\rm MAD}=0.23$~dex, also obtained targeting the RGB stars of the galaxy. The small deviations we found, again of the order of 0.1~dex, can be attributed to the different spectral resolution between the data and implemented technique to obtain the metallicity values (namely by directly fitting the available Fe lines). 

We further analysed our sample by considering each pointing separately. 
In Fig.~\ref{fig:met1}, we show the metallicity distribution per field, for clarity only plotting values obtained with the \citeauthor{Starkenburg2010} calibration.
Visual inspection of the MDFs could give the impression of a decrease in the presence of metal-rich stars from F1 to F3 fields. Therefore, we ran a two-sample two-sided Kolmogorov-Smirnov test, comparing separately F2 and F3 against F1, which is the largest sample and the one with the widest range of metallicities. We found that it is likely that the samples are drawn from the same parent distribution (p-values higher than 0.05), therefore the lack of metal-rich stars in F2 and F3 is due to the lower statistics.
This is consistent with IC~1613 constantly forming stars in a spatially homogeneous manner \citep[see][]{Skillman2014}, at least out to $R_{\rm e}$, which is how far our data extend. 

We found that the median [Fe/H] tends to decrease very slightly from the central to the outer field, as reported in Table~\ref{table:met} for both calibrations used (see also the right panel of Fig.~\ref{fig:met1}). 
Thus, we investigated the presence of a radial metallicity gradient performing a Gaussian process regression (GPR) analysis using a Gaussian kernel together with a noise component to account for the intrinsic metallicity scatter of the data. Details of this type of analysis have been described extensively in \citet{Taibi2022}. For both calibrations we obtained a value of the gradient $\nabla_{\rm [Fe/H]}$ that is consistent with zero within the uncertainties (see Fig.~\ref{fig:met1} and Table~\ref{table:met}), meaning that the bulk of the data, with [Fe/H] values of $\sim-1.0$~dex, do not show significant spatial variation. 
This result is consistent with the conclusions of \citet{Taibi2022}, who found that LG dwarf galaxies, except for those likely to have undergone past mergers, have similar gradient values.

The CaT method only applies to RGB stars, while we have mentioned that \textsc{spexxy} provides spectral parameters, including [Fe/H], for all stars in our sample. However, we have shown in the Appendix~\ref{apx:sanity-checks} that it has limitations in providing correct estimates. Nevertheless, we were able to show qualitatively that the RGB stars should be at least 0.2~dex more metal-poor than the MS stars, as expected from the SFH of IC~1613 \citep{Skillman2014} and spectroscopic measurements of young supergiants \citep{Bresolin2007,Tautvaisiene2007,Berger2018}. 

\section{Discussion}
\label{sec:discussion}

In the previous sections, we have provided the results of the kinematic and chemical analysis of the MUSE data of IC~1613. 
The general picture we get is that both the gaseous and stellar components are rotating, with the evolved stars appearing to decouple from the motion of the HI mainly due to the asymmetric drift. The MS stars, on the other hand, appear to have maintained kinematics similar to that of the neutral gas. The velocity dispersion profile decreases with radius for all tracers, although the central MS value may be affected by the presence of binaries or the expansion motion of the central hole, which may bias it towards a higher value. The lack of neutral gas in the central parts is also the probable cause of the sudden decrease in the inner HI velocity dispersion. Nevertheless, there seems to be a correlation between the velocity dispersion of the different tracers and their age, with values increasing as we move from the gas towards the evolved stars at all radii. At the same time, the evolved stars seem to have a lower rotation support than the HI and MS stars.

In the following, we examine these age-kinematic relations in detail, discussing their implications for the formation mechanisms of IC~1613.
At the same time, we address the fact that the inclination of this galaxy is rather uncertain, proposing an alternative way of calculating it using only stellar tracers of different ages. Although we do not obtain conclusive results, we find the multi-tracer analysis to be a promising way to resolve this issue.

\subsection{Age-kinematics relations}
\label{sec:AVR}

\begin{figure*}
	\centering
	\includegraphics[width=\textwidth]{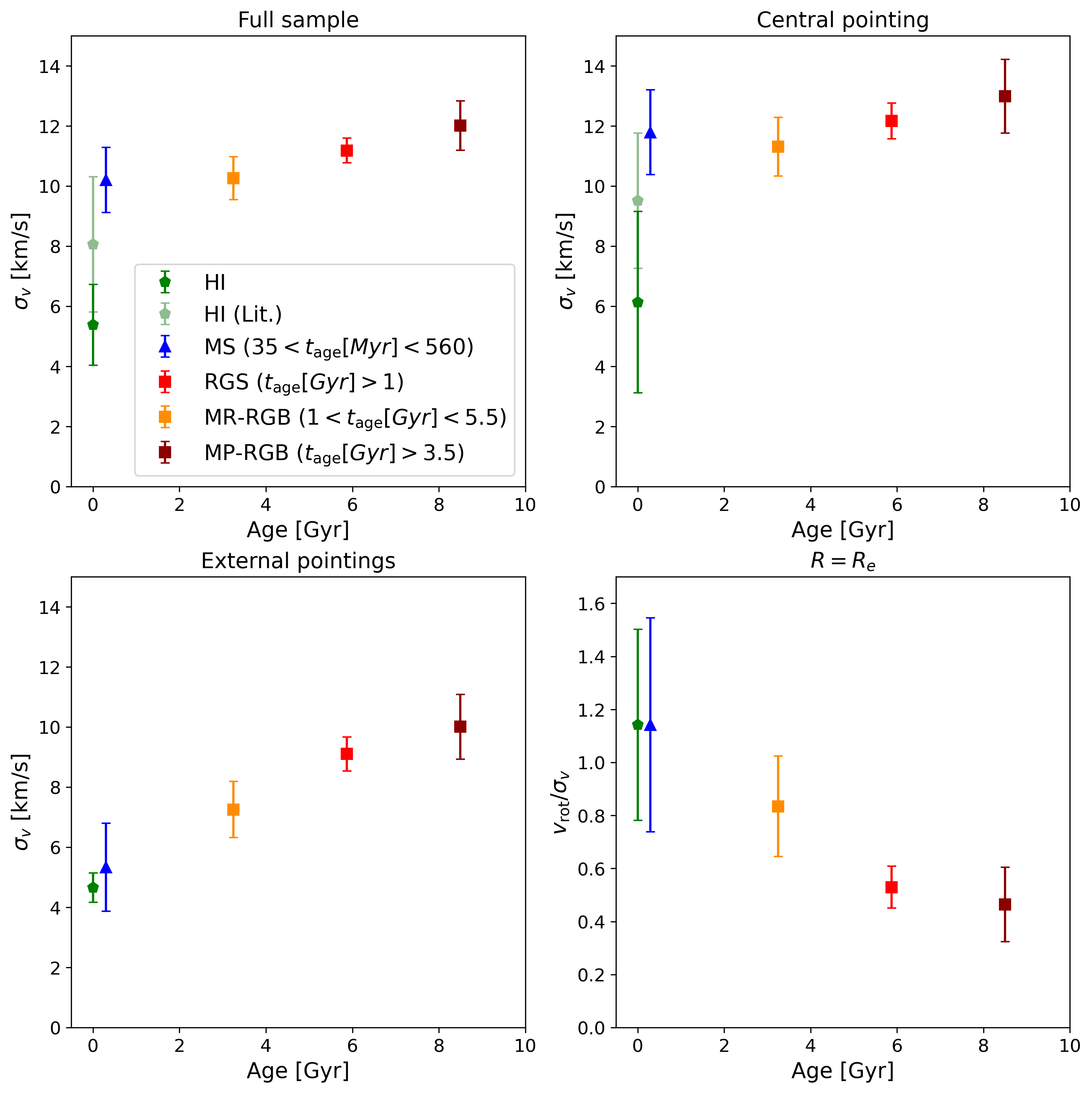}
	\caption{Age-kinematics relations of different tracers. From left to right, from top to bottom, are reported the $\sigma_{\rm v}$ (first three panels) and $v_{\rm rot}/\sigma_{\rm v}$ (last panel) for the different tracers analysed in this work, as a function of their average age (covered ranges are reported in the legend). Markers as in Fig.\ref{fig:kin_rotcurve}, to which we add as orange and maroon squares the values for the MR and MP subsamples of RGB stars, respectively.}
	\label{fig:ADR}
\end{figure*}

In Sect.~\ref{subsec:kin-fields} and \ref{subsec:kin-spType}, we have shown how the different stellar tracers inspected exhibit different kinematic behaviour. Here we make more explicit the connection of these changes with stellar age, also taking into account information recovered from the chemical analysis. In particular, as shown in Fig.~\ref{fig:ADR}, we inspect the variation of the velocity dispersion and of the rotation-to-dispersion support ratio as a function of age. 

In the first three panels of Fig.~\ref{fig:ADR} (from left to right, from top to bottom), we show the age-velocity dispersion variation of the different kinematic tracers. We note that each point has an age value that is only the average of the probable age range covered by the individual tracers. To provide a finer sampling of the age axis, we added two additional subsamples selected from the RGB stars with metallicity measurements. We used their median ${\rm [Fe/H]}$ value (see Table~\ref{table:met}) to divide them into a metal-poor (MP) and a metal-rich (MR) subsample. According to the age-metallicity relation obtained from the SFH of IC~1613 \citep{Skillman2014}, they cover an age range between $1<t_{\rm age, MP}[{\rm Gyr}]<5.5$ and $3.5<t_{\rm age, MR}[{\rm Gyr}]<13.5$, respectively. In all three panels, the $\sigma_{\rm v}$ values of each tracer are reported in Table~\ref{table:multi2}, while for the HI we took the average of $\sigma_{\rm v, HI}$ over the entire radial range (top left panel), the average for $R<3$~arcmin (top right panel) and the average for $R>3$~arcmin (bottom left panel). The average of the $\sigma_{\rm v, HI}$ profile measured for other galaxies in the literature is also shown for comparison.

From the figure, it can be seen that the velocity dispersion tends to increase as we move towards higher ages, both when considering each tracer in its entirety as well as when considering the central and external pointings separately.
In particular, we see that in the top two panels the stellar $\sigma_{\rm v}$ is higher than $\sigma_{\rm v, HI}$, but increases only slightly towards older ages, being consistent with a constant trend when considering the errors. In the bottom-left panel, however, we see a clear linear increase, with the $\sigma_{\rm v}$ of the younger tracers being significantly lower than $\sigma_{\rm v, MP}$.
Considering the presence of the central HI hole, whose causes may have affected not only the kinematics of the HI itself but also that of the MS stars, the external pointings probably provide a more accurate picture of the age-velocity dispersion relation of IC~1613.
In the bottom-right panel of Fig.~\ref{fig:ADR}, we show instead the ratio of rotational to dispersion support (without accounting for the inclination) for both stars and gas as a function of age. The HI value is calculated at the effective radius $R_e$, while the stellar values are the average of the external F2 and F3 pointings (see again Table~\ref{table:multi2}). Again, a linear trend is evident, with $v_{\rm rot}/\sigma_{\rm v}$ decreasing with age, with values between $\sim 0.5 - 1.2$. 

Such age-kinematics trends are in good agreement with those of two similarly bright, gas-rich and isolated LG dwarf galaxies, WLM and NGC~6822. Both systems show a $v_{\rm rot}/\sigma_{\rm v}$ that decreases with age and a $\sigma_{\rm v}$ that grows with age \citep{Leaman2012,Belland2020}\footnote{\citet{Belland2020} found such trends in [Fe/H], which also in the case of NGC~6822 are a good proxy for age.}. 
In particular, an age-velocity dispersion relation (AVR) has also been observed at higher luminosities for the LMC, M33, M31 and the MW, while the satellite dSphs of the MW seem to show a constant trend with age (e.g. \citealp{Leaman2017}, but also \citealp{Dorman2015,Quirk2019,Quirk2022}; and \citealp[e.g.][for AVR studies outside the LG]{Poci2019,Shetty2020,Pessa2023}). 
The AVR has historically been interpreted by assuming that stars are born from a gas that is always in a dynamically cold state, with the increase in velocity dispersion with age due to various scattering mechanisms (e.g. giant molecular clouds or spiral arms) that secularly increase the orbital energy \citep{Spitzer1951,Spitzer1953,Carlberg1985,Kokubo+Ida1992}. 
Alternatively, the AVR could be interpreted by assuming that stars are born from an interstellar medium that becomes less turbulent over time due to a decreased gas fraction and less gravitational instability in the gas disc \citep[e.g.][]{Bournaud2009,Forbes2012}. 
The analysis of \citet{Leaman2017} favours this latter scenario, in which a latent dynamical heating also seems to be at play moving towards higher galactic masses, while environmental effects may have affected the MW satellite systems. Although the coarse age resolution of our sample prevents us from performing a more quantitative analysis of the AVR of IC~1613, our results seem to support the conclusions of \citet{Leaman2017}. 

A similar argument could be made inspecting the variation of the rotation support with age. We have seen that the $v_{\rm rot}/\sigma_{\rm v} (R_e)$ of IC~1613 decrease with age, similarly to WLM and NGC~6822. However, few other systems in the LG have kinematic studies covering the youngest stars, with the majority focusing on the more accessible red giant population. In this context, \citet{Wheeler2017} showed with an homogeneous kinematic analysis that in general the gas-rich systems of the LG are not strong rotators in their evolved stellar component (i.e.~they have $v_{\rm rot}/\sigma_{\rm v} \lesssim 2$). These results appear valid regardless of the environment, in agreement with recent results found for dwarf galaxies residing in voids \citep{deLosReyes2023}. The average rotation-to-dispersion support of the evolved stars in IC~1613 is $v_{\rm rot}/\sigma_{\rm v} (R_e)\sim 0.5$, which remains $\lesssim 2$ even taking into account the uncertainties over the inclination value (see Sect.~\ref{sec:incl+mdyn}). 
This is in agreement with the results of \citet{Wheeler2017} who concluded, comparing with zoom-in cosmological simulations of isolated dwarf galaxies, that dwarf galaxies probably form as puffy dispersion-supported systems, rather than with a cold kinematics. Considering that IC~1613 evolved mostly in isolation, the age trends observed in $v_{\rm rot}/\sigma_{\rm v}$ and $\sigma_{\rm v}$ appear to be intrinsic features of this system, the interpretation of which favours the \citet{Leaman2017} and \citet{Wheeler2017} scenarios. 

The age-kinematic trends of IC~1613 are also in broad agreement with observations of galaxies at different redshift ($z$), where it can be seen that in general $\sigma_{\rm v}$ increases, while $v_{\rm rot}/\sigma_{\rm v}$ decreases, as a function of $z$ \citep[e.g.][]{Wisnioski2015,Wisnioski2019,Ubler2019}. These studies, however, barely cover the mass regime of dwarf galaxies, while recent results seem to question whether high redshift galaxies are dispersion-supported, with a $v_{\rm rot}/\sigma_{\rm v}$ similar to that of local spirals \citep{Rizzo2020,Rizzo2021,Fraternali2021,Lelli2021}. Such systems are more massive than local dwarf galaxies, and at present only a few facilities such as the James Webb space telescope, can help to understand if rotation-supported dwarfs were common in the early Universe \citep[e.g.][]{deGraaff2024}. However, considering the kinematic state of local dwarf galaxies in isolation, including gas-poor ones \citep{Taibi2018,Taibi2020}, this possibility seems less likely \citep{Leaman2017,Wheeler2017}.

Our results on IC~1613 show the importance of performing a multi-tracer chemo-kinematic analysis covering several age ranges. Ongoing and future spectroscopic surveys, such as DESI \citep{Levi2013}, WEAVE \citep{Jin2024} and 4MOST \citep{deJong2019}, will provide such an analysis for the nearest dwarf galaxies with large statistical samples. At the distance of IC~1613 and beyond, we have shown here the challenges inherent in obtaining large samples covering a wide-area, for which we should await the arrival of planned or proposed facilities, such as the Mauna Kea Spectroscopic Explorer \citep[MSE][]{MSE2019} and the Wide-field Spectroscopic Telescope \citep[WST][]{Mainieri2024}.

\subsection{Dynamical mass estimation and the role of inclination}
\label{sec:incl+mdyn}

We previously mentioned that the HI inclination value of IC~1613 is a matter of debate. Using low-resolution HI data, \citet{Lake+Skillman1989} reported a value of $i=38^\circ\pm5^\circ$, in very good agreement with the best-fitting value from \citet[][see Table~\ref{table:1}]{Read2016}, which we have adopted. A similar value was also obtained from \citet[][]{Oh2015}, although unconstrained. However, \citet{Read2016} suggested that the true inclination could be as low as $i=15^\circ$, which allowed to bring the galaxy in agreement with model expectations. In fact, the HI circular velocity curve of IC~1613 grows very slowly, even compared to expectations of a cored dark matter density profile and assuming the most favourable value of low inclination. This appears to be the consequence of depressions and holes in the HI field, with the stellar field probably having been affected by them too, as we saw in Sects.~\ref{subsec:kin-fields} and \ref{subsec:kin-spType}.
For \citet{Collins+Read2022} an inclination value of $\sim20^\circ$ would bring the galaxy in agreement with expectations from the baryonic Tully-Fisher relation (BTFR). We note that such a low inclination is at odds with \citet{Lake+Skillman1989}, who argue that a higher inclination value is to be preferred on the basis that the external isophotes of the HI (assumed as a thin disk) agree well in shape and orientation with those of the stellar component, as is also the case between the spatial and kinematic P.A. of the HI. Indeed from the stellar photometric axis ratio we would obtain an inclination of $i={\rm arccos}(b/a)={\rm arccos}(1-\epsilon)\approx37^\circ$.
If the galaxy was seen almost face-on, it would be difficult to explain such features.

Here we attempt to derive the inclination angle entirely from the stellar component, using the RGS and MS stars as independent dynamical tracers, whose properties have been reported in Sect.~\ref{subsec:kin-spType}. The idea is quite simple: since both tracers must share the same circular velocity, we can equalise the formula for the asymmetric drift correction for both of them and solve for the inclination. In practice, we equalise the right-hand side of the following equation: $v_{\rm circ}^2=v_{\rm rot}^2/{\rm sin}^2(i) + \sigma_{\rm D}^2$, where $\sigma_{\rm D}$ is the asymmetric drift correction term necessary to remove the random motion component that suppresses the rotation curve. Another implicit assumption is that both tracers also share the same inclination angle. 
The $\sigma_{\rm D}$ term could be rather complex, as it takes into account the shape of the velocity ellipsoid and the tracer density profile \citep[see e.g.][]{Weijmans2008}. 

As a first attempt, we follow the formalism of \citet{Read2016} which primarily assumes a kinematic tracer with an exponential surface brightness profile and a constant observed velocity dispersion profile. This reduces the above equation to: $v_{\rm circ}^2 \simeq v_{\rm rot}^2/{\rm sin}^2(i) + \sigma_{\rm v}^2 R/R_{\rm d}$, where $R_{\rm d}$  is the exponential scale radius. We further assume that both tracers share the same scale radius (as $R_{\rm d} = R_{\rm e}/1.68$). To solve for the inclination, we draw $10^5$ random values from the distributions of the rotation and velocity dispersion values, assuming Gaussian errors. We find then an inclination of $i =36^\circ \pm 11^\circ$. Such value is dominated by the observational errors, in particular of the MS stars.

We also attempted a more sophisticated approach following, for instance, the prescription of \citet[][using their Eqs. 6-12 and following considerations in their Appendix A2]{Weijmans2008}. In particular, assuming that the galaxy is axisymmetric, that the rotation curve flattens at large radii and that the potential is spherical, the asymmetric drift correction term becomes:
\[
\sigma_{\rm D}^2= - \sigma_{\rm R}^2 \left [ \frac{\partial {\rm ln} \Sigma}{\partial {\rm ln}R} + \frac{\partial {\rm ln}\sigma_{\rm R}^2}{\partial {\rm ln} R} + \frac{R^2}{2R_{\rm mod}^2} + \frac{\kappa R^2}{\kappa(2R_{\rm mod}^2 - R^2) + R^2} \right ]
\]
where the first term between parenthesis is the slope of the surface brightness profile, again assumed to be exponential, while the last two terms are connected to the shape of the velocity ellipsoid and the shape of the rotation curve. The former is modelled with a parameter $ \kappa \in [ 0;1]$, while the latter assumes that the rotation curve follows a power-law model where the curve becomes flat at large radii. The radius at which the rotation curve starts to flatten is called core radius $R_{\rm c}$ and enters in the equation as $R_{\rm mod}^2 = R^2 + R_{\rm c}^2$. We calculate $R_{\rm c}$ from the HI rotation curve ($7.2\pm1.9$~arcmin) and assume it is the same also for the stellar tracers. While calculating $i$, we also randomise over these two terms assuming a uniform distribution for $\kappa$ and a Gaussian one for $R_{\rm c}$. Nevertheless, their contribution to the final value resulted to be minor. The radial dispersion is instead parametrised as $\sigma_{\rm R} (R) = \sigma_0 + \sigma_1 {\rm exp}(-R_{\rm mod}/R_{\rm d})$, which can be obtained from the observed $\sigma_{\rm v}$ following Eq.~(11) of \citet{Weijmans2008}. Such equation depends on $i$, but also on $\kappa$ and $R_{\rm c}$. Therefore, in order to find a parametric solution of $\sigma_{\rm R}$, we also randomise over $i$ that will act as a prior distribution, which we assume uniform over $ u \in [ 0;1]$, with $i={\rm arccos}(u)$. 
Compared to the simpler approach described above, we find here a slightly lower inclination value, $i =32^\circ \pm 11^\circ$, but a comparable uncertainty. 

Both recovered inclinations lie in the middle of the literature values, with the error bars preventing one from favouring a high \citep{Lake+Skillman1989} or low value \citep{Read2016,Collins+Read2022}. It seems though that a value as low as $15^\circ$ is less favoured. If an inclination of $\sim35^\circ$ will be preferred, this will confirm IC~1613 as an outlier of the BTFR. Such behaviour seems unusual for a galaxy with a baryonic mass of $M_b=9.5\pm0.5\times10^7 M_\odot$ \citep{Read2016}, while it seems to resemble that of gas-rich ultra-diffuse galaxies found in the field (though an order of magnitude more massive; see \citealp{Mancera-Pina2019}). We defer to a future dynamical study the task of better quantifying the value of inclination, where all available kinematic tracers, be they gas and stars, are simultaneously taken into account \citep[see e.g.][for WLM]{Leung2021}. 

We can now estimate IC~1613's dynamical mass, based on the stellar rotational velocities, corrected for the galaxy's inclination, and the asymmetric drift; therefore turned into circular velocities $v_{\rm circ}$. For the inclination, we stick to the last calculated value of $32^\circ \pm 11^\circ$, while for simplicity we assume again the asymmetric drift correction from \citet{Read2016} in order to estimate $v_{\rm circ}$. 
We perform this estimate within the 3D de-projected half-light radius $r_{1/2}\simeq (4/3) R_{\rm e}$ to make a direct comparison with values from mass estimators for dispersion supported systems. 
We take the average of the observed values of $v_{\rm rot}$ and $\sigma_{\rm v}$ between F2 and F3, assuming that both the rotation and dispersion profiles are slowly varying up to the half-light radius. The circular velocity is then $v_{\rm circ} (r_{1/2})=15\pm2$~km\,s$^{-1}$. 
The dynamical mass $M_{\rm dyn} (r_{1/2})=v_{\rm circ}^2 r_{1/2}/G$ is then $1.1\pm0.3\times10^8 M_\odot$, while the corresponding mass-to-light ratio within the half-light radius would be $M_{\rm dyn}/L_V(r_{1/2})=2.2\pm0.6 \, M_\odot/L_\odot$.

For comparison, the \citet{Wolf2010} mass-estimator for dispersion-supported systems, that is $M_{1/2} = 3 G^{-1} \sigma_{\rm v}^{2} r_{1/2} \approx 4 G^{-1} \sigma_{\rm v}^{2} R_{\rm e}$, where \textit{G} is the gravitational constant, yields $M_{1/2}=1.0\pm0.2\times 10^8 M_\odot$ within the same radius. This is within the previous results accounting for inclination, but also within the value reported by \citet{Kirby2014}. We note that this mass estimator is valid as long as the system is spherically symmetric, without rotation and with a flat l.o.s.~velocity dispersion profile. To some degree, all these conditions are violated for IC~1613. 

\section{Summary and future work}
\label{sec:conclusions}

In this paper, we present results from the chemo-kinematic analysis of the stellar component of the isolated dwarf irregular galaxy IC~1613.
The analysis is based on a new set of spectroscopic data obtained with 3 pointings of the VLT/MUSE instrument. We extracted $\sim 2000$ sources, from which we separated the stellar objects for their subsequent spectral classification and analysis. 

The quality of the dataset allowed for a generally accurate spectral classification (i.e., with $T_{\rm eff}$ determined to better than 500~K) of 808 stars. We found a majority of K-type stars, and representatives from all the other types (from O-type to M-type stars). The classification included the identification of a sample of 24 probable Be stars, together with a sample of 14 probable C stars.

For all types except the Be stars, the l.o.s.~velocities have been determined with a pixel-by-pixel analysis, resulting in measurements with a mean precision of $\delta_{\rm v} = 7$~km\,s$^{-1}$ and signal-to-noise ratio S/N$_{\rm C} \sim10$. For the Be stars we have determined the velocities by cross-correlation with dedicated templates.
From this sample we identified 746 probable members (i.e., having $P_M>0.95$), which show a significant rotation pattern. When modelled as a linear velocity gradient this yields $k=1.1\pm0.3$~km\,s$^{-1}$\,arcmin$^{-1}$, along the HI kinematic major axis (with P.~A.~$=72^\circ$). This is the first time that rotation for the stellar component is detected with high significance for this galaxy. 

We further analysed the probable members looking at the kinematic properties of each individual field and as a function of stellar types tracing different broad age ranges. The individual fields analysis confirmed the presence of a rotation signal, and a velocity dispersion profile that decreases with radius. On average, the stellar component shows higher velocity dispersion and a slower rotation than the HI gas; this is a manifestation of the asymmetric drift, as demonstrated by the fact that the kinematics of the two tracers comes in good agreement when applying an asymmetric drift correction. The kinematic analysis of the different selected stellar populations showed a coherent picture with the analysis of the main sample, with the MS stars following more closely the kinematics of the HI gas than the evolved red giant stars.

The chemical analysis was conducted on the sub-sample of RGB stars exploiting the wavelength region of the Ca~II triplet lines. We obtained a median ${\rm [Fe/H]} = -1.06$~dex, with scatter values of $\sigma_{\rm MAD} = 0.29$~dex and $\sigma_{\rm intrinsic} = 0.26$~dex, in good agreement with previous results reported by \citet{Kirby2013}. The average [Fe/H] is well within the rms scatter of the stellar luminosity-metallicity relation found for the LG dwarf galaxies by the same authors. The spatial distribution of the [Fe/H] values showed no significant signs of a radial metallicity gradient, compatible with results from other LG dwarf galaxies \citep{Leaman2013,Taibi2022}. 

Examining the kinematics of stars as a function of broad age ranges, we find that the velocity dispersion increases as a function of age, with the behaviour being very clear in the outermost pointings, while the rotation-to-velocity dispersion support decreases. On timescales of $< 1$~Gyr, the stellar kinematics still follow very closely that of the neutral gas, while the two components decouple on longer timescales. We interpret these results as a consequence of stars being born from a less turbulent gas over time. On the other hand, with the dynamical mass estimation we provide a new value $i=32^\circ \pm 11^\circ$ of the inclination angle of IC~1613. To this end, we use stellar tracers with different observed kinematics that must be consistent with the same underlying gravitational potential.

In future efforts, the data presented here could be used to perform a dedicated analysis of the massive stars in our sample using the \textsc{FASTWIND} stellar atmosphere code \citep{Puls2005}, which will allow us to obtain reliable stellar parameters from their spectra. 
We also aim to improve the identification process of the M-type stars in our sample, and therefore to constrain the spectroscopic C/M ratio. This will tell us more of the chemical properties of the intermediate-age stellar population of IC~1613. 
This MUSE data-set could also be used to analyse the ionised gas medium using available emission lines, like $H_\alpha$ or nebular oxygen lines in order to get an insight on the kinematics of the shell-like gas structure visible in field F1 and compare it with that of the youngest stars.
Finally, we aim to perform a dedicated dynamical study where all available kinematic tracers, be they gas and stars, are simultaneously taken into account in order to firmly establish the (dark) matter distribution of IC~1613.

\begin{acknowledgements}
We wish to thank the anonymous referee for the constructive comments that helped to improve the manuscript.
We also thank M.-R.~Cioni, L.~R.~Patrick, J.~Read, F.~Lelli, and I.~Trujillo for useful discussions and comments at various stage of this project.
ST acknowledges funding of a Leibniz-Junior Research Group (PI: M.~Pawlowski; project number J94/2020) via the Leibniz Competition.
GB and CG acknowledge support from the Agencia Estatal de Investigaci{\'o}n del Ministerio de Ciencia en Innovaci{\'o}n (AEI-MICIN) and the European Regional Development Fund (ERDF) under Grant Number PID2020-118778GB-I00/10.13039/50110001103 and the AEI under Grant Number CEX2019-000920-S.
GI acknowledges financial support from the European Research Council (ERC) through the ERC Consolidator Grant DEMOBLACK, under contract No.~770017.
SK acknowledges funding from UKRI in the form of a Future Leaders Fellowship (grant no.~MR/T022868/1).
PEMP acknowledges the support from the Dutch Research Council (NWO) through the Veni grant VI.Veni.222.364.
This research has made use of NASA’s Astrophysics Data System, VizieR catalogue access tool (CDS, Strasbourg, France, DOI: 10.26093/cds/vizier), and extensive use of Python3.8 \citep{Python3}, including iPython \citep[v8.12,][]{ipython}, Numpy \citep[v1.24,][]{NumPy-Array}, Scipy \citep[v1.10,][]{SciPy-NMeth}, Matplotlib \citep[v3.7,][]{Matplotlib}, Astropy \citep[v5.2,][]{Astropy} and Scikit-learn \citep[v1.2,][]{scikit-learn} packages.
\end{acknowledgements}

%
%

\bibliographystyle{bibtex/aa} 
\bibliography{bibtex/ic1613.bib} 


\begin{appendix}

\section{Calibration of MUSE catalogues}
\label{subsec:photometry}

The raw catalogues obtained from the MUSE \textit{VRI} pseudo-images contain the spatial and photometric information of all extracted sources, which is needed for the kinematic and chemical analysis we conducted. However, they needed astrometric and photometric calibration. For this purpose, we used the publicly available HST/WFPC2 catalogues from \cite{Holtzman2006}, which corresponded to the F1 and F3 fields, and a proprietary Subaru/SuprimeCam catalogue for the F2 field, kindly provided to us by M.~Monelli (IAC). 

We note that the field-of-view (FoV) of the Subaru catalogue ($34\arcmin \times 27\arcmin$) is wide enough to cover all the three MUSE fields at once. However, it lacked an astrometric solution at first, and it showed severe crowding toward the galaxy's centre, where F1 is placed. The HST catalogues, on the other hand, did not show any of these problems and we used them as a baseline during the astrometric and photometric calibration processes. 

We downloaded the PSF-extracted HST catalogues with transformed magnitudes in the Johnson's system\footnote{Catalogues can be found at \url{http://astronomy.nmsu.edu/holtz/archival/ic1613/html/ic1613.html}}. The covered photometric bands were \textit{V} and \textit{I}, with the addition of the \textit{B}-band for the central HST pointing. The Subaru catalogue covered instead the Johnson's \textit{B} and \textit{V} bands. These are deep photometric catalogues, reaching down to the horizontal branch (at $V\sim25$), which means that they cover a much wider photometric range than that of our MUSE data.
Before proceeding, we performed a pre-cleaning, selecting all sources marked as stellar in the HST catalogues, while retaining those targets in the Subaru catalogue having DAOPHOT parameters sharpness (SHARP) between $-0.5$ and $0.5$, and a goodness-of-fit (CHI) $<1$.

The astrometric solution and photometric calibration were obtained using the suite of codes \textit{CataXcorr} and \textit{CataComb}, kindly provided to us by P.~Montegriffo and M.~Bellazzini (INAF-OAS). 
We started with the astrometric registration of the Subaru catalogue. Due to its large FoV compared to that of the HST catalogues, we first performed a pre-astrometric registration using the stars in common with a PanSTARRS catalogue \citep{PanSTARRS2016} generated within the FoV of the Subaru catalogue. We note that the PanSTARRS catalogue was shallower that the Subaru one, reaching only a magnitude lower the tip of the RGB (at $I\sim 21$). We then used the external HST field catalogue to refine the astrometric solution, improving its accuracy to better than 0.05\arcsec. 

The photometric bands of the MUSE catalogues we were interested in calibrating were the \textit{V} and \textit{I} bands. We recall that the raw catalogues extracted from the data cubes were made in the \textit{VRI} bands. The \textit{B}-band was not extracted because it is not completely covered by the wavelength range of MUSE. 
Since the HST catalogues covered the BVI and VI bands (central and outer pointing, respectively), while the Subaru only covered the BV bands, as a compromise, we decided to add the \textit{I}-band to the Subaru catalogue and calibrate the MUSE catalogues in \textit{VI}.

We inferred the Subaru \textit{I}-band using the common targets with the central HST catalogue (which covered the \textit{B}-band). We applied the following linear equation: $I_{\rm HST} - V_{\rm HST}=c_{I}\,(B_{\rm HST}-V_{\rm HST})+ZP_I$. 
We obtained a colour term $c_I=-0.99$ and a zero-point $ZP_I=-0.08$, which allowed to calculate the $I_{\rm Sub}$ magnitudes, whose agreement with the $I_{\rm HST}$ of the common targets was good, showing a median absolute deviation (MAD) scatter of 0.08~mags. However, comparing the $(V-I)$ colours of the common sources, we found a difference of $\sim0.2$~mags driven by the bluer objects. This implied the introduction of a colour bias for the probable MS stars of the F2 field, as can be appreciated in Fig.~\ref{fig:cmd_instr}, whose calibration was based on the Subaru catalogue. However, their calibrated $(V-I)$ resulted safely below 0.5~mags (i.e. the colour limit of the RGB).

Finally, we used the mentioned catalogues to astrometrically and photometrically calibrate the corresponding MUSE fields -- the HST ones with the F1 and F3 fields, while the Subaru one with the F2 field. For the photometric calibration of the MUSE pointings using the reference catalogues, we applied the following equations:
$V_{\rm MUSE} - V_{\rm ref} = c_V\,(V_{\rm ref}-I_{\rm ref}) + ZP_V$, and $I_{\rm MUSE} - I_{\rm ref} = c_I\,(V_{\rm ref}-I_{\rm ref}) + ZP_I$.
The accuracy of the astrometric and photometric solutions resulted on average 0.08\arcsec and 0.05~mags, respectively. 

\section{Consistency checks}
\label{sec:apx}

\subsection{Verify SPEXXY general performance}
\label{apx:sanity-checks}

We verified the performance of \textsc{spexxy} in recovering l.o.s.~velocities and associated errors. We recall that \textsc{spexxy} performs a full spectral fit to the observed spectra using interpolated templates generated from the PHOENIX library of high-resolution synthetic spectra \citep{Husser2013}. It requires a configuration file specifying initial values for velocity, effective temperature, surface gravity, metallicity and [$\alpha$/M] from which to start the fit. We obtained the initial stellar parameters from a set of isochrones \citep{Girardi2000,Bressan2012} with ${\rm [Fe/H]}=-1$~dex (around the mean metallicity of IC~1613) by performing a linear regression between the observed colours and magnitudes and the theoretical $T_{\rm eff}$ and ${\rm log}(g)$. We also set an initial velocity guess of $-230$~km\,s$^{-1}$ (i.e. approximately IC~1613 systemic velocity).
For the spectral fitting, we used the available PHOENIX templates, convolved with the MUSE line spread function, with solar scale chemical composition, thus fixing ${\rm [\alpha/M]}=0$~dex. We let the code find the best fit of all other parameters (i.e., $T_{\rm eff}$, ${\rm log}(g)$ and ${\rm [Fe/H]}$, together with the l.o.s.~velocity).
We also masked certain spectral regions to avoid possible emission lines due to the interstellar medium and residuals due to sky line subtraction (such as the persistent residual due to the [OI] line at 5577~\AA).
	
To test for possible systematic velocity shifts we ran the code on several noiseless templates from the PHOENIX library of giant stars at different $T_{\rm eff}$ (from 2500~K to 15000~K) at fixed ${\rm [Fe/H]}=-1.5$~dex. Templates were shifted at several l.o.s.~velocities, from $-500$~km\,s$^{-1}$ to $50$~km\,s$^{-1}$ at step of 50~km\,s$^{-1}$. As initial guess for the velocity fitting, we set again $-230$~km\,s$^{-1}$.
We found that in general \textsc{spexxy} correctly recovers the assigned l.o.s.~velocities (mean and scatter of the difference between input and output velocities being $-0.06\pm0.56$~km\,s$^{-1}$), although we noticed some failure around the extremes (i.e. for $-500$~km\,s$^{-1}$ and $50$~km\,s$^{-1}$) with velocities variations of up to 50\%. This occurred in only 3\% of the cases that are those differing the most from the initial guess for the velocity.
We recall that based on the Besan\c{c}on model we expect almost no contaminants with true velocities $\sim-500$~km\,s$^{-1}$, while the variation at $50$~km\,s$^{-1}$ is small enough to keep such possible contaminants away from the bulk velocity of IC~1613. 

Regarding the other parameters, $T_{\rm eff}$ values were also well recovered, except for some scatter of $\sim2000$~K at temperatures $>10000$~K; ${\rm log}(g)$ was recovered with a significant scatter ($>2$~dex) for the hottest stars and those with $T_{\rm eff}\sim6500$~K (the latter a minority in our main sample); ${\rm [Fe/H]}$ was also well recovered, but with a large scatter ($>1$~dex) for templates with extreme velocities. Again, this is mostly due to the large difference between the initial guess and the true velocity of the stellar template, which leads \textsc{spexxy} to an incorrect parameter estimate.
Therefore, \textsc{spexxy} generally recovers accurate velocities and spectral parameters.

To test whether \textsc{spexxy} provides the correct uncertainties, we developed a series of Monte Carlo tests in which we generated mock spectra at different S/N, starting from noiseless templates of different spectral types and injecting noise directly from the error spectra associated to the observed ones. However, we found that the flux uncertainty associated with the observed spectra was underestimated by an average factor of 1.4 with respect to the measured noise on the observed flux. This was found systematically for all spectra, regardless of their spectral type or S/N. It is most likely a result of the resampling into a regularly sampled data cube during data reduction. This introduces covariances between pixels that are neglected by the pipeline.
In principle, this factor is not a problem, as the uncertainties of \textsc{spexxy} are calculated from the residuals between the input spectrum and the best-fit template, but we had to take this into account when generating the mock spectra.

Since we have two main stellar populations in our observed sample, RGS and MS stars, we generated mock spectra representative of these two sub-samples. For the RGS, we selected stars with $T_{\rm eff}\sim5000$~K, with a difference between observed values found by \textsc{spexxy} and ULySS of no more than 500~K. From their best-fitting templates and scaled noise spectra, we generated 250 mock spectra in bins of S/N$_{\rm CaT}$ around $[3.5, 5, 10, 15, 25]$. The same was done for the MS stars, but for stars with $T_{\rm eff}\sim10000$~K, generating mock spectra at S/N$_{\rm 550}$ around $[5, 10, 20, 50]$. Results are reported in Table~\ref{table:mock-vel}.

We found that for the RGS stars their velocities and associated errors are well recovered, with deviations between the mean of the errors and the measurement scatter within $0.5$~km\,s$^{-1}$. The effective temperature was also well recovered, although the errors were as much as $\sim50$~K lower than the measurement scatter. On the other hand, ${\rm log}(g)$ and ${\rm [Fe/H]}$ were relatively more uncertain, with errors underestimated between 0.1 and 0.2~dex (higher deviations at lower S/N), leading to an offset on the recovered values of up to 0.1~dex. 

For the MS stars we recovered higher deviations from the input values than the RGS case. In particular, velocity errors at S/N$_{550}=5$ were underestimated by more than $15$~km\,s$^{-1}$, while at the highest S/N the deviation was less than 1~km\,s$^{-1}$. In the main text we reported that we made a quality cut in our sample by excluding stars with observed velocity errors $>25$~km\,s$^{-1}$. This cut mainly excluded MS stars with the lowest S/N, while ensuring that we kept those with well-estimated velocity errors within 10\%. For the spectral parameters, the associated errors were generally more underestimated ($>10\%$) than in the RGS case. 

Therefore, it can be concluded that \textsc{spexxy} is generally effective in accurately and precisely recovering stellar velocities. However, it struggles to correctly recover stellar parameters, with ${\rm log}(g)$ being the most uncertain parameter in our mock tests. This issue is primarily related to the low resolution of the MUSE data rather than a problem with the fitting routine.
It is also important to note that our tests only explored a limited region of the parameter space. To further evaluate the performance of \textsc{spexxy} in recovering ${\rm [Fe/H]}$ values, we conducted an additional test by comparing the values obtained from \textsc{spexxy} with those obtained for the RGB stars using the CaT method (see details in Sect.~\ref{sec:metallicity}). 

We found a significant deviation between the distributions of ${\rm [Fe/H]}$ values from the two methods of $\sim0.4$~dex calculated between the medians, with those obtained with \textsc{spexxy} being metal richer.
The recovered deviation is significantly larger than expected based on the mock tests (see again Table~\ref{table:mock-vel}) and did not show a dependence with S/N. This discrepancy can be partly attributed to the poor recovery of log(\textit{g}) by \textsc{spexxy}, as shown by the mock tests. In fact, we re-run \textsc{spexxy} on our targets while fixing log(\textit{g}) to the initial input values, thus reducing the number of free parameters to fit. As a result, the offset to the CaT-based values decreased to $\sim0.2$~dex. The observed difference remains significant, but the distributions now have comparable standard deviations. 

Fixing the value of log(\textit{g}) during the \textsc{spexxy} fitting did not significantly change the ${\rm [Fe/H]}$ distribution of the MS stars. This could be due to the lower sensitivity of the MS stars to log(\textit{g}) variations when recovering metallicity values. The results of the mock tests show that the recovered values are quite accurate (for S/N$>10$), but with slightly underestimated errors. 

Assuming that fixing the log(\textit{g}) yields metallicity values accurate to within a constant shift in the RGB case, we can make a relative comparison between the ${\rm [Fe/H]}$ distribution of RGB and MS stars using \textsc{spexxy} outputs. We further refined the selection of MS stars by keeping the probable kinematic members with $T_{\rm eff}<15000$~K (i.e. up to the grid limit of the PHOENIX library) and S/N$_{550}>15$ to ensure a similar sample quality as for the RGB. This reduced the MS sample to 30 stars. Their average metallicity was comparable to that of the RGB sample, but with a larger associated standard deviation. Taking into account that the average value for the RGB may be overestimated by 0.2~dex, we can conclude that the MS stars are expected to be more metal-rich than the RGB. This difference may increase if we take into account the ${\rm [\alpha/Fe]}$ for the two populations, which we set to zero during the \textsc{spexxy} fitting.
The fact that the MS stars are more metal-rich than the RGB would be expected \citep{Skillman2014}, since also spectroscopic measurements of young blue and red supergiants from the literature suggest an average ${\rm [Fe/H]}=-0.7$~dex \citep{Bresolin2007,Tautvaisiene2007,Garcia2014,Berger2018}, while the ionised HII medium may be slightly metal-rich (by $\sim0.1$~dex, \citealp[e.g.][]{Bresolin2007}). 

\begin{table*}
	\caption{Results from the Monte Carlo tests.}             
	\label{table:mock-vel}      
	\centering          
	\begin{tabular}{ccccc|rrrr}
		\hline\hline
		S/N & $V_{\rm rad,0} $ & $T_{\rm eff,0}$ & ${\rm log}(g)_0$ & ${\rm [Fe/H]}_0$ & $\frac{(V_{\rm rad}-V_{\rm rad, 0})}{\delta_{\rm V_{\rm rad}}}$ & $\frac{(T_{\rm eff}-T_{\rm eff,0})}{\delta_{\rm T_{\rm eff}}}$ & $\frac{({\rm log}(g)-{\rm log}(g)_0)}{\delta_{\rm {\rm log}(g)}}$ & $\frac{({\rm [Fe/H]}-{\rm [Fe/H]}_0)}{\delta_{\rm {\rm [Fe/H]}}}$ \\ 
		(pxl$^{-1}$) & (km\,s$^{-1}$) & (K) & (dex) & (dex) &  &  &  & \\
            \hline
		3.5 & $-211\pm14$ & $5523\pm457$ & $3.3\pm1.2$ & $-0.2\pm0.4$ & $ 0.0, 1.0$ & $ 0.0, 1.2$ & $ 0.1, 1.2$ & $ 0.0, 1.1$ \\
		5   & $-247\pm9 $ & $5065\pm289$ & $2.7\pm0.9$ & $-0.8\pm0.3$ & $ 0.1, 0.9$ & $ 0.0, 1.1$ & $-0.1, 1.2$ & $ 0.2, 1.2$ \\
		10  & $-235\pm5 $ & $5499\pm145$ & $2.3\pm0.5$ & $-0.3\pm0.2$ & $ 0.0, 1.0$ & $ 0.3, 1.1$ & $ 0.2, 1.2$ & $ 0.2, 1.1$ \\
		15  & $-223\pm4 $ & $4851\pm109$ & $1.3\pm0.4$ & $-0.7\pm0.2$ & $ 0.0, 0.9$ & $ 0.1, 1.3$ & $-0.1, 1.2$ & $ 0.2, 1.3$ \\
		25  & $-234\pm2 $ & $4965\pm64 $ & $2.3\pm0.3$ & $-0.6\pm0.1$ & $ 0.1, 1.0$ & $ 0.0, 1.1$ & $ 0.0, 1.3$ & $ 0.0, 1.0$ \\
		\hline
		5   & $-249\pm29$ & $9090\pm1077$ & $3.9\pm0.9$ & $ 1.0\pm2.2$ & $-0.2, 1.5$ & $-0.7, 1.2$ & $-0.3, 1.1$ & $-0.7, 0.5$ \\
		10  & $-223\pm13$ & $7206\pm187$  & $3.5\pm0.5$ & $-0.8\pm0.3$ & $ 0.0, 1.1$ & $-0.3, 1.2$ & $-0.3, 1.8$ & $ 0.1, 1.4$ \\
		20  & $-228\pm7$  & $7581\pm94$   & $3.0\pm0.2$ & $-1.2\pm0.2$ & $-0.1, 1.1$ & $ 0.0, 1.3$ & $-0.1, 1.5$ & $ 0.1, 1.5$ \\
		50  & $-224\pm3$  & $11213\pm117$ & $2.9\pm0.1$ & $-0.6\pm0.1$ & $ 0.0, 1.2$ & $ 0.0, 1.2$ & $ 0.1, 1.3$ & $ 0.1, 1.2$ \\
		\hline
	\end{tabular}
    \tablefoot{From left to right, the first column indicates the S/N obtained from the continuum around the CaT lines, for the first five rows, and around 5500~\AA, otherwise; the next fours columns are the l.o.s.~velocity, effective temperature, surface gravity and metallicity recovered by \textsc{spexxy} for the selected stars; the last four columns are the mean and standard deviation of the difference between the mock values and the measured ones, scaled for the mock errors of the considered parameters.}
\end{table*}

\subsection{Comparison between SPEXXY and ULySS}
\label{apx:spexxy-vs-ulyss}

\begin{figure*}
    \centering
    \includegraphics[width=0.45\textwidth]{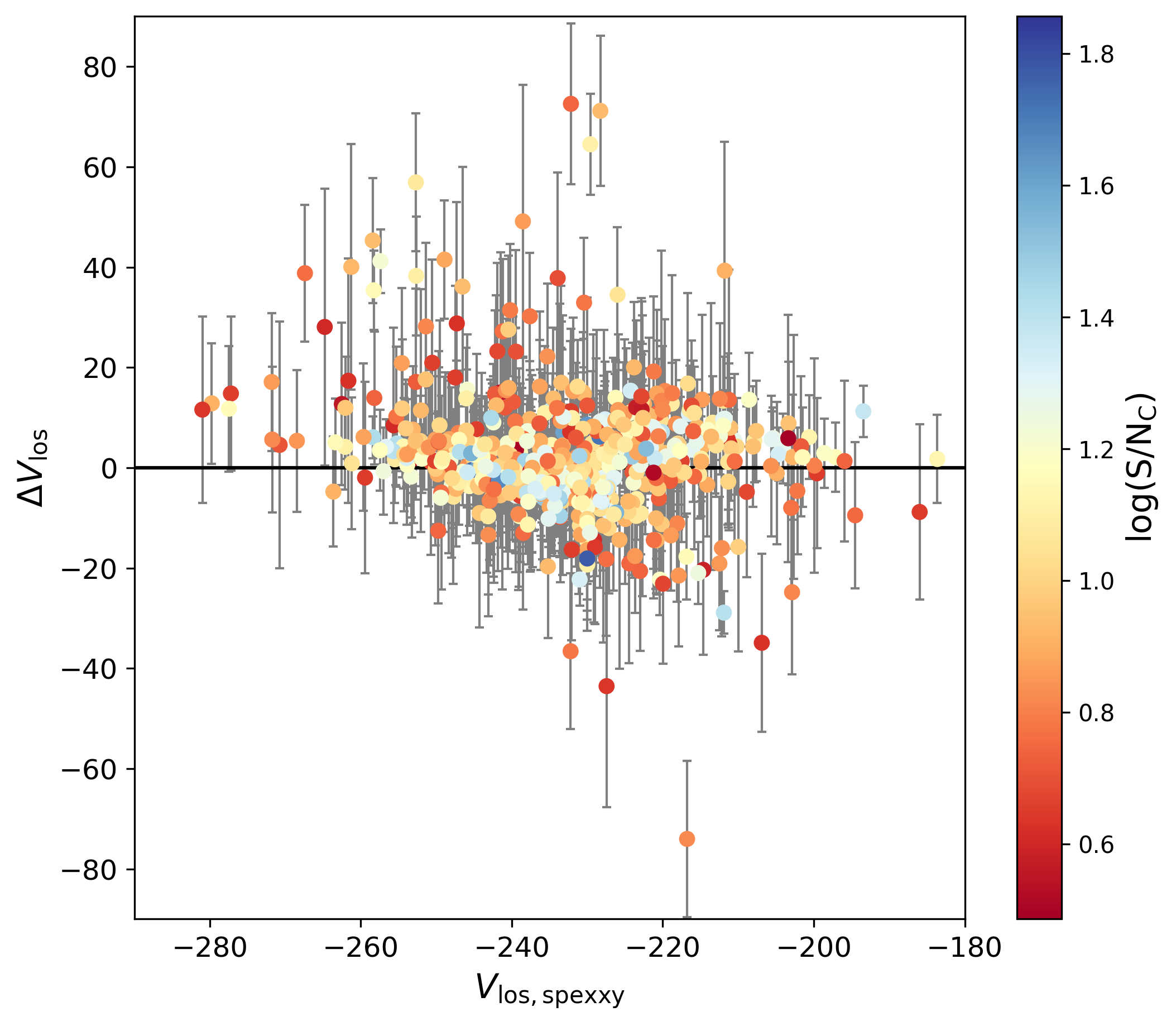}
    \includegraphics[width=0.45\textwidth]{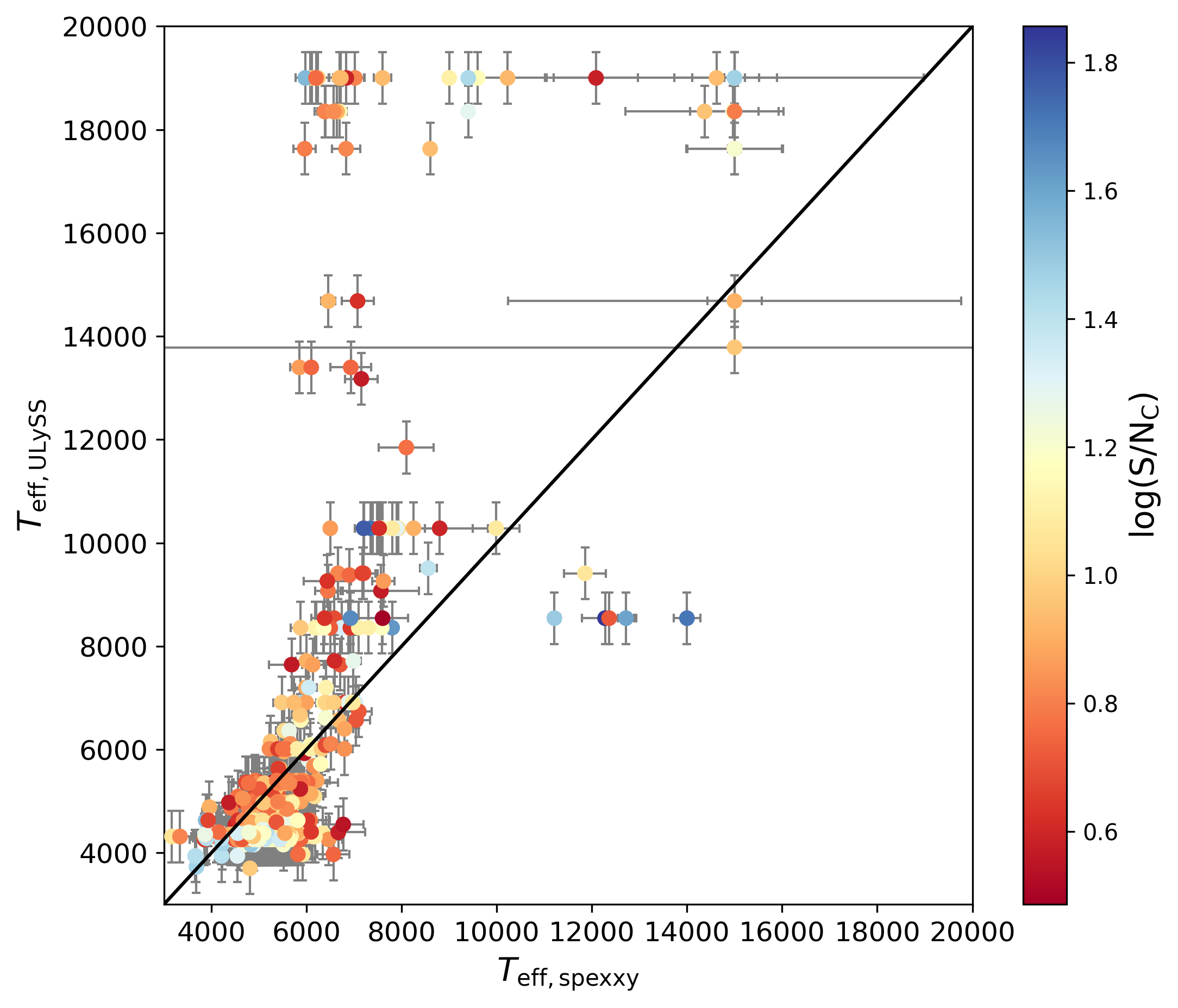}
    \caption{Comparison between \textsc{spexxy} and ULySS. \textit{Left:} Velocity differences for pair measurements obtained with the two codes. The solid black line represents the zero velocity offset. Filled circles are colour-coded according to their S/N$_{\rm C}$, while the grey error-bars indicate the combined uncertainty, that is $\sqrt{\delta^2_{\rm rad, Spexxy}+\delta^2_{\rm rad, ULySS}}$. \textit{Right:} direct comparison between the effective temperature values obtained with the two codes; the solid black line is for reference only.}
    \label{fig:spexxy_vs_ulyss}
\end{figure*}

In our analysis, we used \textsc{spexxy} and ULySS in a complementary way: the former for l.o.s.~velocity measurements, while the latter for spectral classification. However, since both codes provide similar outputs, we compared their performances in recovering the l.o.s.~velocity and $T_{\rm eff}$ from the observed spectra. 
	
We considered all targets in the sample used for the kinematic analysis (see Sect.~\ref{sec:kinematics}).
The l.o.s.~velocity errors for the ULySS measurements were obtained from Monte Carlo simulations using several templates of giant stars at different intervals of temperature and S/N. We made this choice because the formal errors returned by ULySS were generally overestimated, without following any clear trend with the S/N or spectral type. For the Monte Carlo simulations we chose seven templates from the MIUSCAT library (see Sect.\ref{subsec:ULySS}), with $T_{\rm eff}$ ranging from 3500~K to 20000~K. The simulations were performed by adding Poisson noise to each template to obtain simulated spectra at increasing S/N intervals, with values ranging from 4 to 30. We repeated each process 100 times. For each template, the recovered velocity scatter as a function of S/N was fitted by an exponential law. The ULySS measurements were then divided into effective temperature bins according to the templates considered, and the velocity errors assigned using the corresponding exponential fit. 
	
First we compared the velocity differences for pair measurements obtained with ULySS and \textsc{spexxy}. We found that $\sim75$\% of the measured velocities are in agreement within the combined uncertainties, obtained as $\sqrt{\delta^2_{\rm rad, Spexxy}+\delta^2_{\rm rad, ULySS}}$. The comparison is shown in the left panel of Fig.~\ref{fig:spexxy_vs_ulyss} where for clarity we reported only those measurements with a combined error $<30$~km\,s$^{-1}$. Nevertheless, the difference between the ULySS and \textsc{spexxy} measurements showed an average positive shift of $\sim 5$~km\,s$^{-1}$. What is more, a higher shift of $\sim 10$~km\,s$^{-1}$ is observed for those objects with a S/N$_{\rm C}<10$. This was somehow expected for the UlySS measurements, since the Monte Carlo tests showed average velocity shifts of up to 20 km\,s$^{-1}$ for templates with effective temperatures of $\sim4500$~K and greater than $\sim10000$~K having simulated $S/N<10$. On the other hand, we have already showed in the previous section the capacity of \textsc{spexxy} to correctly recover l.o.s.~velocities, even at the lowest S/N for red giant stars. This is a very important point because most of the stars in our dataset are RGS in a low S/N regime ($\lesssim10$), where \textsc{spexxy} showed its strength in the recovery of l.o.s.~velocities.
	
The comparison of effective temperatures, as shown in the right-hand panel of Fig.~\ref{fig:spexxy_vs_ulyss}, gave good results for $T_{\rm eff}<7500$~K, although the \textsc{spexxy} values tend to be systematically $\sim500$~K higher than the ULySS values. 
At higher $T_{\rm eff}$, \textsc{spexxy} values started to deviate significantly. This is mainly due to the fact that \textsc{spexxy} measurements are based on the PHOENIX library, which is not optimized for very hot stars. In the ULySS case, instead, for each spectrum the stored temperature value is that of the best fitting template, which was chosen by eye among the most likely templates. We recall that this procedure had an accuracy of up to $\sim500$~K, for a quality PCL-flag between 3 and 4.

\subsection{Biases in the recovery of kinematic properties}
\label{apx:kin-bias}

We checked for the presence of possible biases that could alter the recovery of the kinematic properties of the stellar samples, in particular when inspecting the properties of the RGS and MS stars. On one hand, the region approximately separating the MS from the RGS, $0.4\lesssim(V-I)\lesssim0.8$ (see Fig.~\ref{fig:CMD-Teff}), contains the instability strip of IC~1613, where it is expected to find mainly classical Cepheids with short-term variability (i.e., with log$(P[{\rm days}])<0.5$ and $t_{\rm age}<0.5$~Gyr; \citealp[see e.g.][]{Udalski2001,Bernard2010}). Since we observe them in random phases, we expect any differences to be averaged out. It is also true that, especially for Cepheids, hydrogen or metallic lines can lead to different velocities \citep[see, e.g.,][]{Vinko1998}. Nonetheless, we find no significant differences when calculating the average velocities and velocity dispersion including or excluding stars in that color range. 
Similar considerations can be made for the Be-star candidates in our sample, which should also be short-period variables \citep[e.g.,][]{Porter+Rivinius2003,Rivinius2013}, although in this case it is mostly the intensity of the emission lines that is affected. We repeated the same test finding values compatible within the errors. 
As for the RGS sample, this contains a small number of C-star candidates, which are long period variables \citep[e.g.,][]{Menzies2015}. However, we have verified that the C-stars have similar properties to the rest of the sample within the errors.

We did not check for the presence of binary stars, which could particularly affect the velocity dispersion values. The low S/N of the individual exposures for each field in our sample and the short period over which they were observed (between a few days and a month), preclude a quantitative analysis. We can, however, make qualitative estimates.
For a sample of mostly RGB stars with high measured velocity dispersion ($>10$~km\,s$^{-1}$), as it is our case, the impact of unresolved binaries is expected to be minimal (i.e. $\sim10-20\%$ of inflation; \citealp[see][]{Minor2010,Minor2013,Spencer2017,Spencer2018,Arroyo-Polonio2023}). However, the case is different when focusing on the MS stars sample. The multiplicity fraction (i.e. the fraction of stars having at least one companion) is generally expected to be $\sim60\%$ for A-types, which increase to $80\%-100\%$ for OB stars \citep[e.g.][]{Moe+DiStefano2017,Offner2023}. In particular, these stars are often found in triples or higher order multiples. For Be stars such figures are less constrained, but mass accretion through binary interactions has been proposed as a possible channel for their formation \citep{Pols1991,deMink2013,Shao+Li2014,Bodensteiner2020}.

Considering that the multiple OBs cover a range of orbital periods, preferably a few days or a few decades \citep{Moe+DiStefano2017}, the stacking process to generate the spectra should already average the short-period ones. For the long-period ones (i.e. log$(P[{\rm days}])>3.5$) we expect velocity variations of $\sim3-10$~km\,s$^{-1}$, which could translate into a velocity dispersion inflation of $\lesssim20\%-30\%$ (following \citealp{Minor2010}, assuming that they behave as short-period low-mass binaries and have intrinsic $\sigma_{v,0}=7-10$~km\,s$^{-1}$). 
This would especially affect the central pointing, where the fraction of OB stars is high ($\sim70\%$). It could thus partly explain why in this field the value of the velocity dispersion of the MS stars is comparable to that of the RGS, while in the other fields it tends to be closer to that of the HI. Another possible source of disturbance for the MS stars in F1 is that they may still be coupled to the motion of the ionised bubble, which is probably expanding \citep[with a typical velocity of $\sim10-20$~km\,s$^{-1}$,][]{Pokhrel2020}. We also do not exclude the presence of possible runaway OB stars that could further impact the kinematics of the sample \citep[e.g.][]{Stoop2024}.


\section{Observing log and examples of extracted spectra}
\label{apx:spectra}

We report here a table with the observing log of our VLT/MUSE observations of IC~1613.
We show as well some examples of the extracted spectra analysed in this work.

\begin{table*}
    \caption{Observing log of the VLT/MUSE observations of IC~1613 under ESO programme 097.B-0373 (PI: G.~Battaglia).}   
    \label{table:obs_log}      
    \centering          
    \begin{tabular}{c c c c c c}    
	\hline\hline
	Field & Position (RA, Dec)  & Date / Hour & Exp. & Airmass & DIMM Seeing \\ 
	& (J2000) & (UT) & (s) & & (arcsec) \\
	\hline           
	F1  &  01:04:49.00, +02:07:16.3  &  2016-08-12 / 09:01  &  1385  &  1.14  &  0.79 \\
	    &  01:04:49.03, +02:07:16.0  &  2016-08-12 / 09:26  &  1385  &  1.17  &  0.68 \\
	    &  01:04:49.00, +02:07:15.6  &  2016-08-13 / 07:57  &  1385  &  1.12  &  0.65 \\
	    &  01:04:48.97, +02:07:15.8  &  2016-08-13 / 08:23  &  1385  &  1.12  &  0.64 \\
	    &  01:04:49.03, +02:07:15.6  &  2016-08-14 / 08:29  &  1385  &  1.12  &  0.67 \\
	    &  01:04:49.00, +02:07:15.9  &  2016-08-14 / 08:54  &  1385  &  1.14  &  0.68 \\
	    &  01:04:48.97, +02:07:16.3  &  2016-09-02 / 03:40  &  1385  &  1.76  &  0.60 \\
	    &  01:04:48.98, +02:07:15.3  &  2016-09-02 / 04:05  &  1385  &  1.56  &  0.62 \\
	\\
	F2  &  01:04:40.69, +02:04:45.2  &  2016-09-28 / 04:00  &  1385  &  1.19  &  0.73 \\
	    &  01:04:40.70, +02:04:46.1  &  2016-09-28 / 04:25  &  1385  &  1.15  &  0.54 \\
	    &  01:04:40.67, +02:04:45.6  &  2016-09-29 / 05:40  &  1385  &  1.13  &  0.52 \\
	    &  01:04:40.73, +02:04:45.9  &  2016-09-29 / 06:05  &  1385  &  1.15  &  0.56 \\
	    &  01:04:40.73, +02:04:45.4  &  2016-09-30 / 03:56  &  1385  &  1.18  &  0.69 \\
	    &  01:04:40.70, +02:04:45.7  &  2016-09-30 / 04:21  &  1385  &  1.14  &  0.71 \\
	    &  01:04:40.72, +02:04:45.2  &  2016-09-30 / 04:52  &  1385  &  1.12  &  0.76 \\
	    &  01:04:40.68, +02:04:45.9  &  2016-09-30 / 05:17  &  1385  &  1.12  &  0.73 \\
	\\
	F3  &  01:04:29.44, +02:03:35.3  &  2016-10-04 / 04:41  &  1385  &  1.12  &  0.67 \\
	    &  01:04:29.39, +02:03:35.4  &  2016-10-04 / 05:06  &  1385  &  1.12  &  0.55 \\
	    &  01:04:29.42, +02:03:35.5  &  2016-10-04 / 05:36  &  1385  &  1.14  &  0.70 \\
	    &  01:04:29.42, +02:03:36.0  &  2016-10-04 / 06:01  &  1385  &  1.17  &  0.83 \\
	    &  01:04:29.41, +02:03:35.1  &  2016-12-21 / 01:16  &  1385  &  1.21  &  0.79 \\
	    &  01:04:29.39, +02:03:35.9  &  2016-12-21 / 01:41  &  1385  &  1.28  &  0.64 \\
	    &  01:04:29.43, +02:03:35.7  &  2016-12-22 / 01:05  &  1385  &  1.20  &  0.92 \\
	    &  01:04:29.44, +02:03:36.0  &  2016-12-22 / 01:30  &  1385  &  1.26  &  0.92 \\ 
	\hline
    \end{tabular}
    \tablefoot{From left to right, column names indicate: the pointing field name; the field center coordinates; observing date and starting time of the scientific exposure; the exposure time in seconds; the starting airmass; the average DIMM seeing during the exposure in arcsec.}
\end{table*}

\begin{figure*}[th!]
    \includegraphics[width=\textwidth]{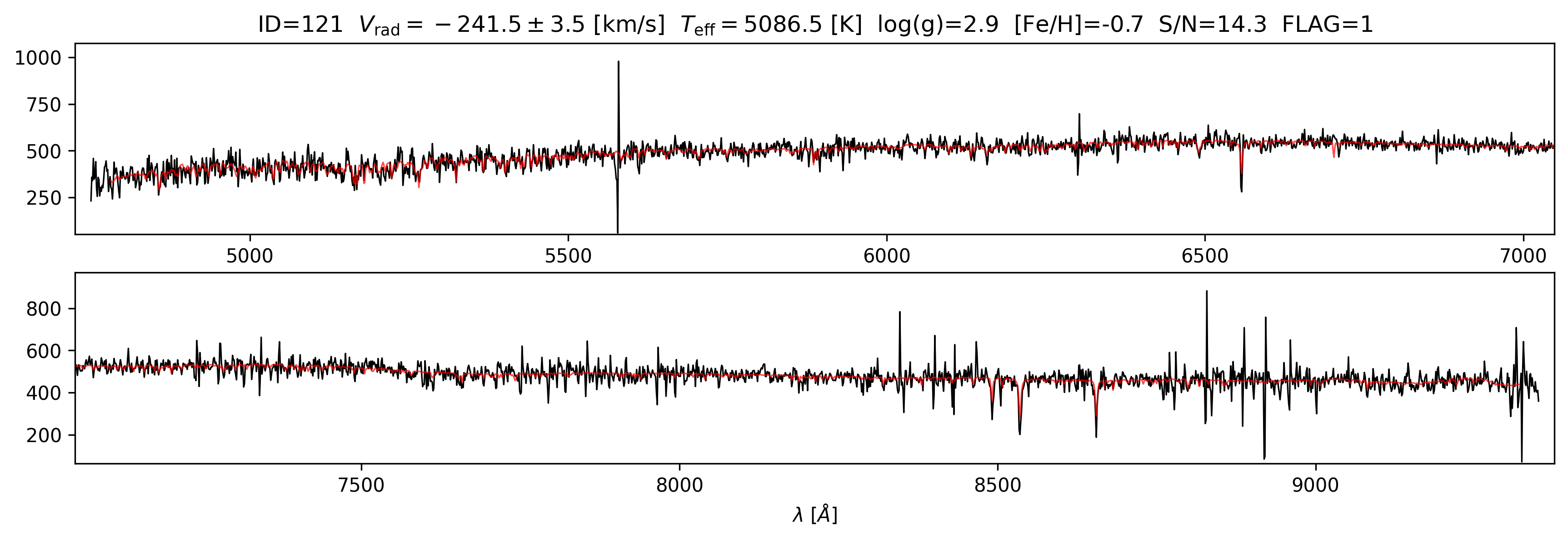}
    \includegraphics[width=\textwidth]{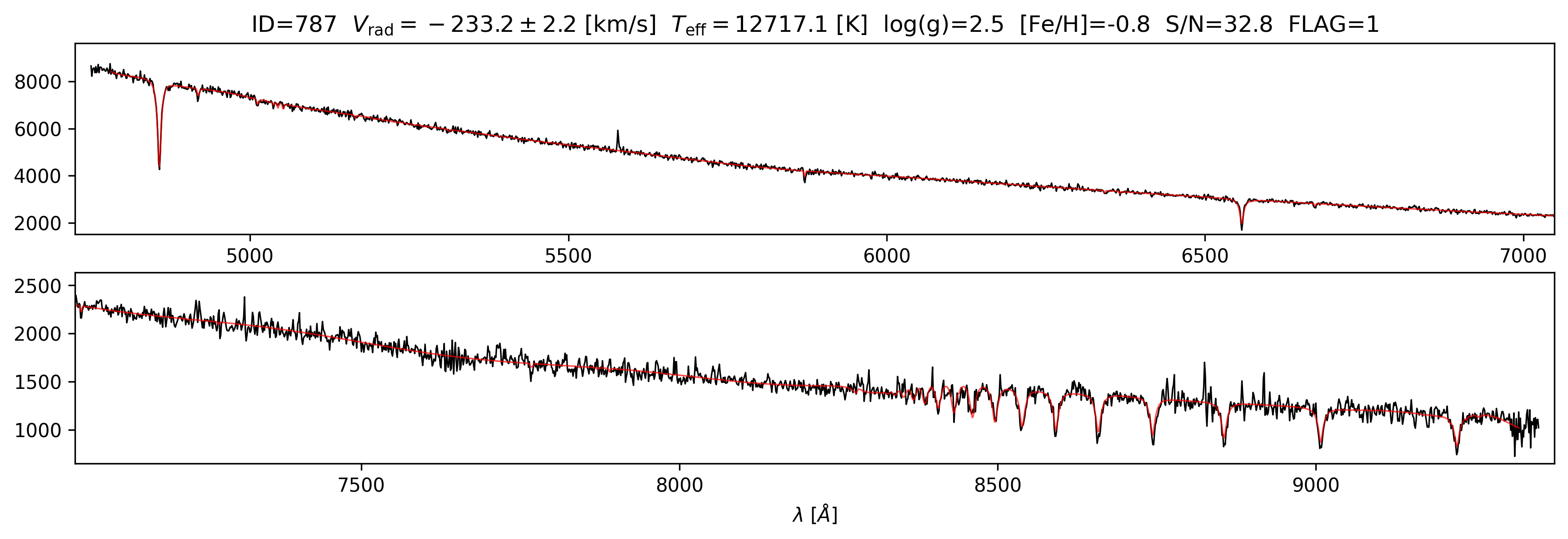}
    \includegraphics[width=\textwidth]{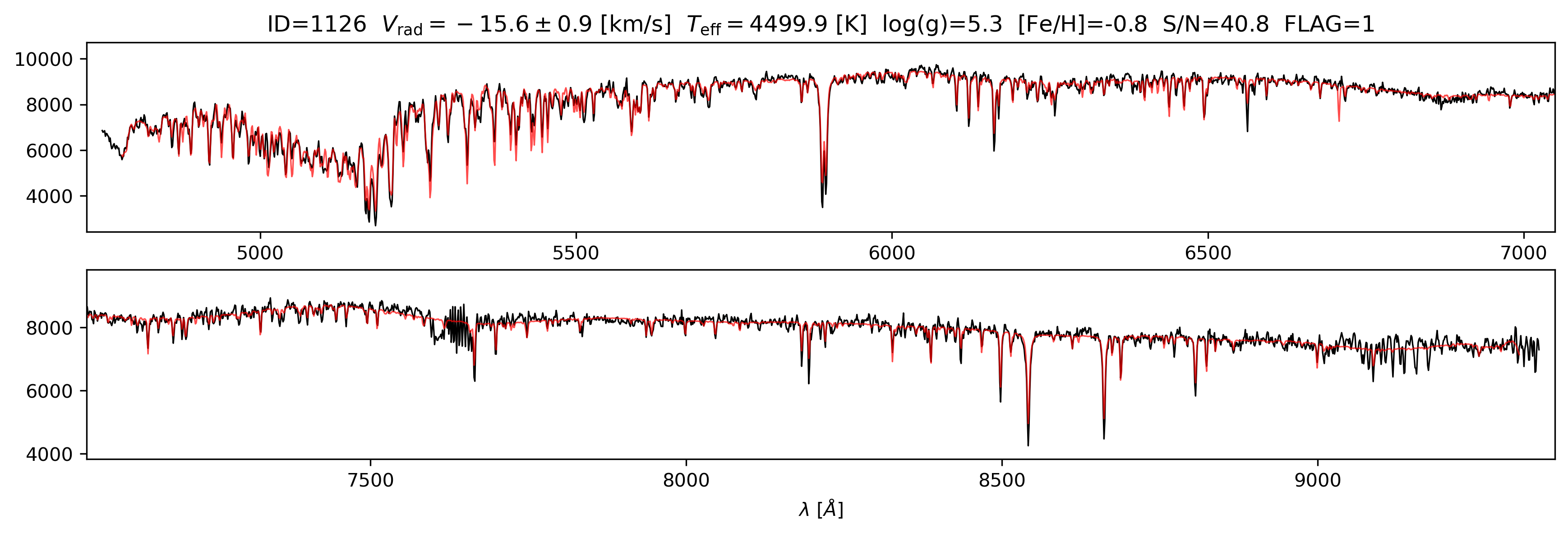}
    \caption{Example of sky-subtracted, wavelength calibrated, extracted spectra from our dataset. In red the best-fitting templates provided by \textsc{spexxy}, with the output spectral parameters (identification, $V_{\rm rad}$, $T_{\rm eff}$, log(\textit{g}), [Fe/H], S/N, binary quality flag) indicated in the labels. From top to bottom examples of: an RGB star; a hot MS star; a foreground star.}
    \label{fig:spectra1}
\end{figure*}

\clearpage

\begin{figure*}[th!]
    \includegraphics[width=\textwidth]{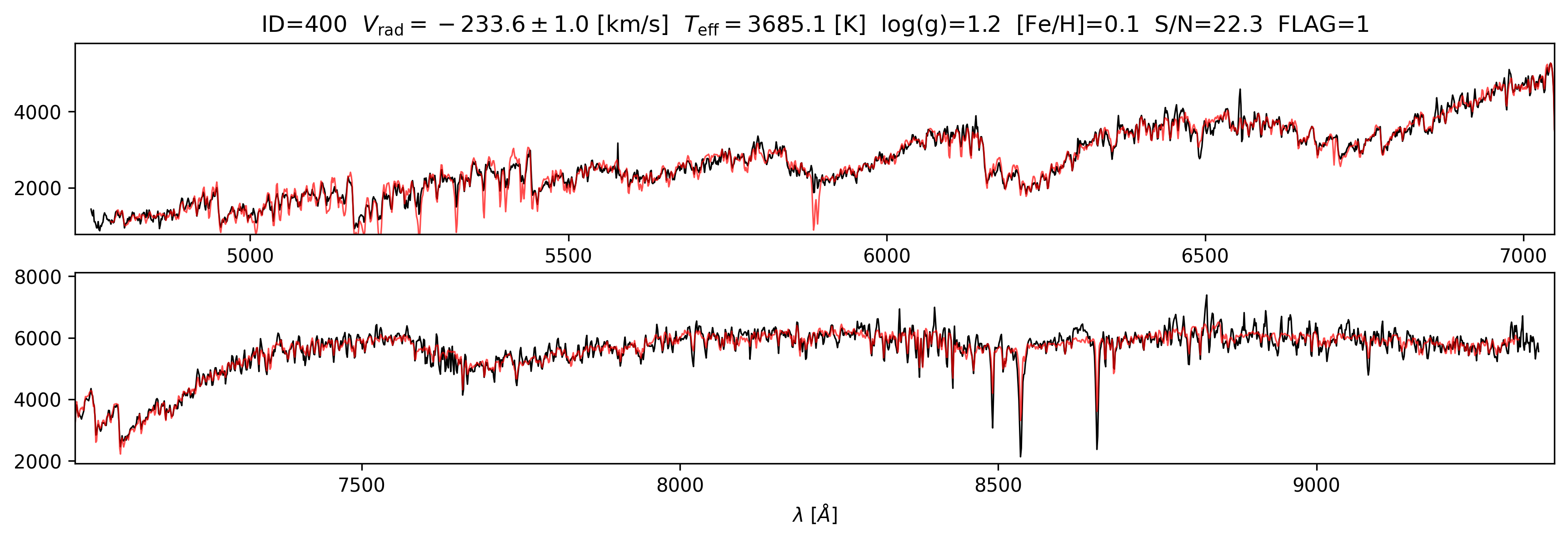}
    \includegraphics[width=\textwidth]{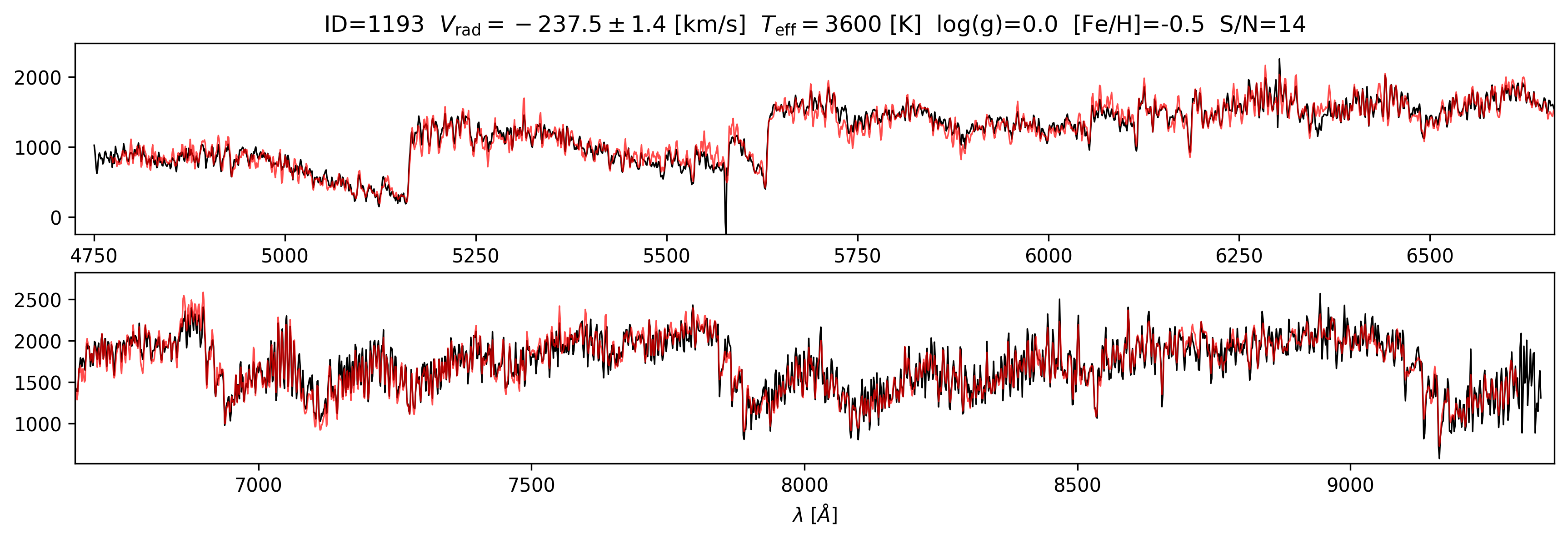}
    \includegraphics[width=\textwidth]{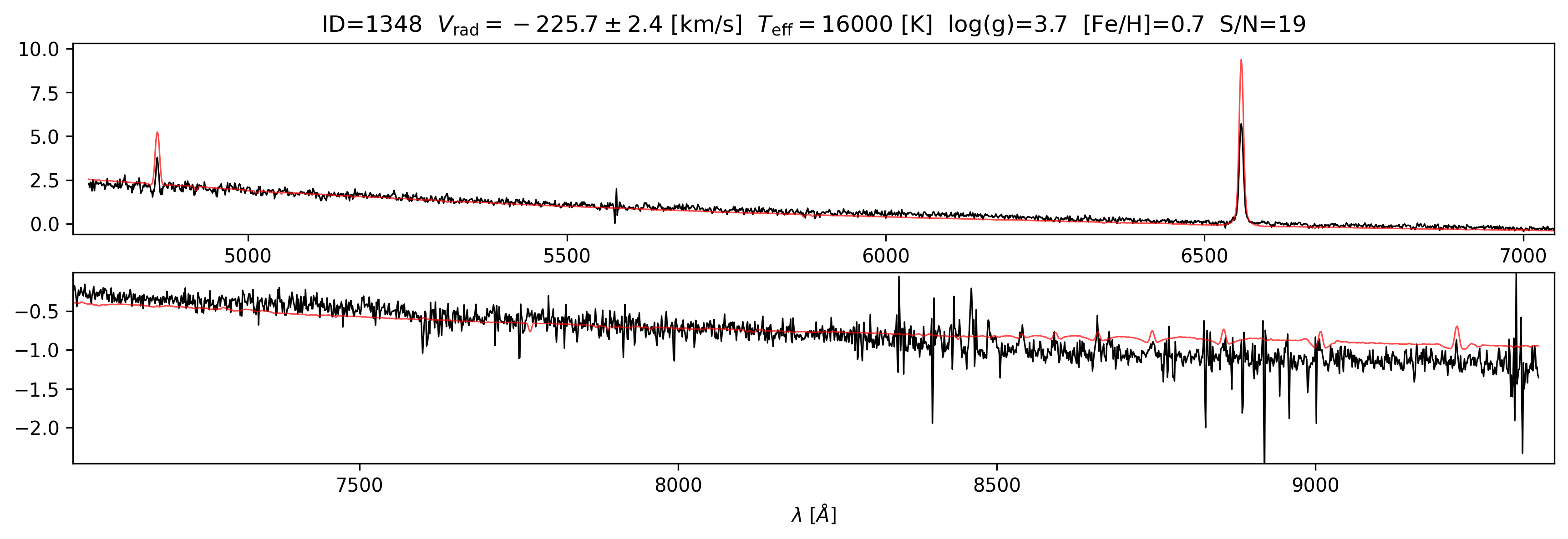}
    \caption{Figure~\ref{fig:spectra1}. Continued. From top to bottom examples of: an M giant star; a C star; a Be star.}
    \label{fig:spectra2}
\end{figure*}

\clearpage
 
\begin{figure*}[th!]
    \includegraphics[width=\textwidth]{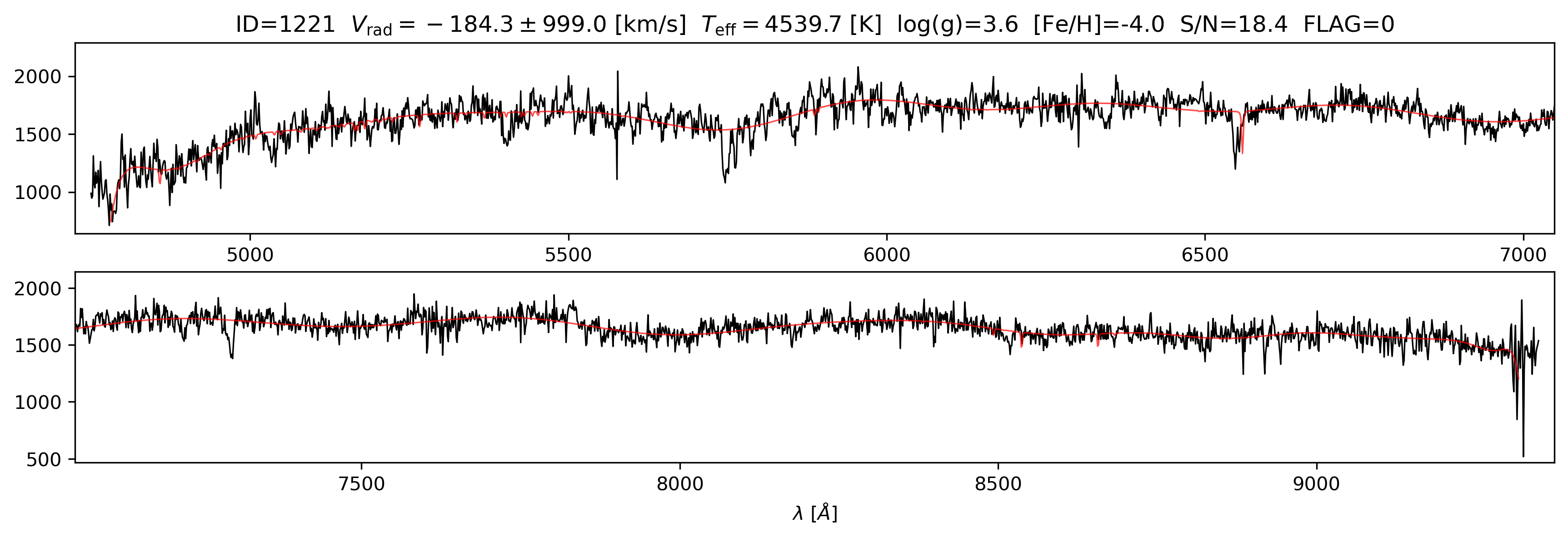}
    \includegraphics[width=\textwidth]{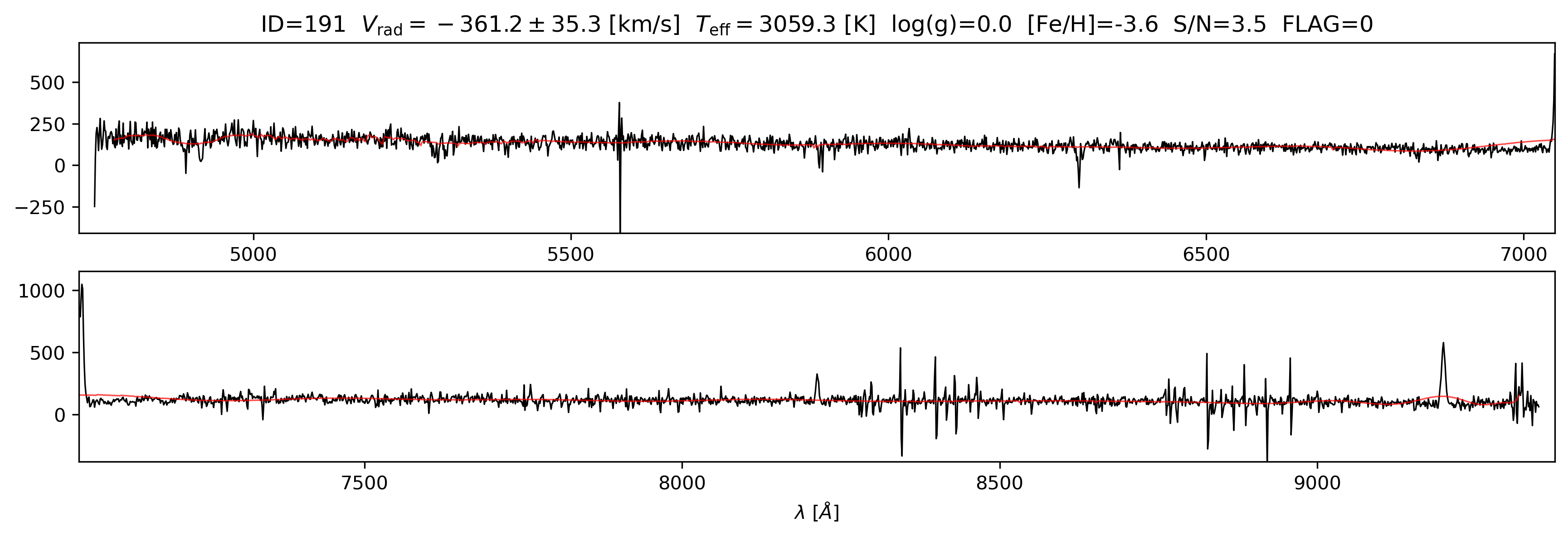}  
    \includegraphics[width=\textwidth]{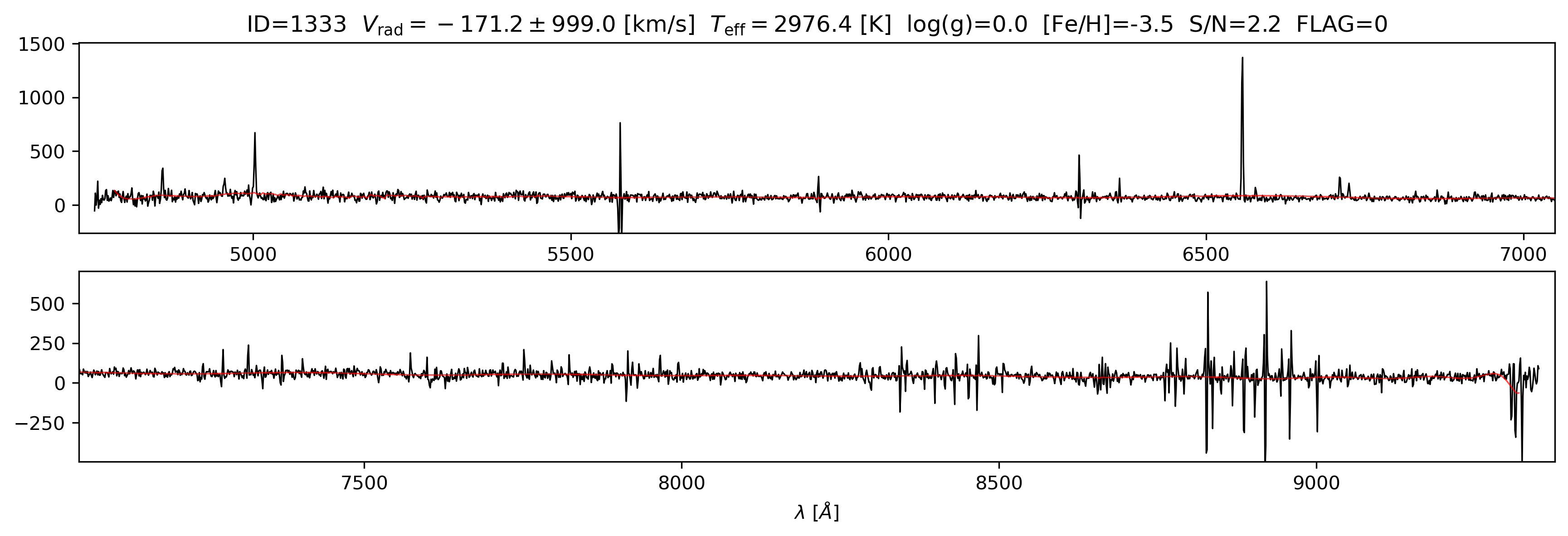}
    \caption{Figure~\ref{fig:spectra1}. Continued. From top to bottom examples of: a background galaxy; a blend with an high redshift emitting-galaxy; an example of ionized gas in IC~1613.}
    \label{fig:spectra3}
\end{figure*}

\clearpage

\end{appendix}


\end{document}